\documentclass{rstransa}

\usepackage{tgpagella}

\usepackage{xcolor}
\graphicspath{{figs/}}

\usepackage{hyperref}
\usepackage{url}         
\usepackage{booktabs}
\usepackage{microtype}
\usepackage[printonlyused,withpage]{acronym}
\usepackage{tabularx}
\usepackage{booktabs}
\usepackage{multirow}
\usepackage{subcaption}
\usepackage{eufrak}
\usepackage{etoolbox}

\usepackage{amssymb,amsfonts,amsmath,amscd}
\usepackage{bm}
\usepackage{mathrsfs}
\usepackage{mathtools}
\usepackage{scalerel}
\allowdisplaybreaks

\usepackage[noabbrev]{cleveref}
\crefname{equation}{}{}

\theoremstyle{definition}
\newtheorem{identity}{Identity}[section]

\usepackage[numbers]{natbib}
\usepackage{tikz}
\usepackage{tikz-3dplot}
\usetikzlibrary{%
  3d,
  calc,
  arrows.meta,
  perspective,
  positioning,
  shapes.arrows,
  decorations.pathmorphing,
  decorations.pathreplacing%
}

\usepackage{xparse}

\let\originalleft\left
\let\originalright\right
\renewcommand{\left}{\mathopen{}\mathclose\bgroup\originalleft}
\renewcommand{\right}{\aftergroup\egroup\originalright}

\NewDocumentCommand\Real{}{ \mathbb{R} }
\NewDocumentCommand\Euclid{}{ \mathbb{E} }

\NewDocumentCommand\bbm{}{ \begin{bmatrix} } %
\NewDocumentCommand\ebm{}{ \end{bmatrix} }   %
\NewDocumentCommand\T{}{\top}                %

\NewDocumentCommand\Vector{m}{ \bm{\mathbf{#1}} }

\NewDocumentCommand\Matrix{m}{ \bm{\mathbf{#1}} }

\NewDocumentCommand\Transpose{m}{ \left.{#1}\right.^\T }

\NewDocumentCommand\Inv{m}{{#1}^{-1}}
\NewDocumentCommand\Trace{m}{ \mathrm{tr}\left(#1\right) }
\NewDocumentCommand\Determinant{m}{ \mathrm{det}\left(#1\right) }
\NewDocumentCommand\Norm{m}{ \left\Vert#1\right\Vert }

\NewDocumentCommand\Zero{}{ \Matrix{0} }

\NewDocumentCommand\Identity{}{ \Matrix{I} }

\NewDocumentCommand\LieGroupGL{m}{ \mathrm{GL}(#1) }

\NewDocumentCommand\LieGroupSO{m}{ \mathrm{SO}(#1) }
\NewDocumentCommand\LieAlgebraSO{m}{ \mathfrak{so}(#1) }

\NewDocumentCommand\LieGroupSE{m}{ \mathrm{SE}(#1) }

\NewDocumentCommand\LieGroupSETwo{m}{ \mathrm{SE_{2}}(#1) }

\NewDocumentCommand\LieGroupGal{m}{ \mathrm{Gal}(#1) }

\NewDocumentCommand\LieGroupSGal{m}{ \mathrm{SGal}(#1) }
\NewDocumentCommand\LieAlgebraSGal{m}{ \mathfrak{sgal}(#1) }

\NewDocumentCommand\LieGroupAdjoint{m}{ \mathrm{Ad}\left(#1\right) }
\NewDocumentCommand\LieAlgebraAdjoint{m}{ \mathrm{ad}\left(#1\right) }

\NewDocumentCommand\LeftJacobianSGal{}{ \Matrix{J}_{l} }
\NewDocumentCommand\RightJacobianSGal{}{ \Matrix{J}_{r} }

\NewDocumentCommand\Wedge{m}{\left.#1\right.^\wedge}
\NewDocumentCommand\Vee{m}{\left.#1\right.^\vee}

\NewDocumentCommand\Matlog{m}{\mathrm{ln}\left(#1\right)}

\NewDocumentCommand\Matexp{m}{\exp\left(#1\right)}

\NewDocumentCommand\Expectation{m}{ \mathbb{E}\left[#1\right] }
\NewDocumentCommand\NormalDistribution{mm}{ \mathcal{N}\left(#1, #2\right) }

\NewDocumentCommand\Estimate{m}{ \hat{#1} }
\NewDocumentCommand\Prior{m}{ \check{#1} }
\NewDocumentCommand\Mean{m}{ \overline{#1} }
\NewDocumentCommand\Observation{m}{ \tilde{#1} }

\NewDocumentCommand\CoordinateFrame{m}{%
  \underrightarrow{ \Matrix{\mathcal{F}}}_{#1} }

\NewDocumentCommand\Defined{}{ \triangleq }

\NewDocumentCommand\<{}{\mspace{1.0mu}}
\NewDocumentCommand\nhs{}{\mspace{-1.5mu}}
\NewDocumentCommand\nds{}{\mspace{-6.0mu}}

\makeatletter
\newcommand{\vast}{\bBigg@{3}}
\newcommand{\Vast}{\bBigg@{4}}
\makeatother

\newcommand{\uw}{\Vector{u}^{\wedge}}
\newcommand{\nuw}{\Vector{\nu}^{\wedge}}
\newcommand{\puw}{\Vector{\rho}^{\wedge}}

\NewDocumentCommand\Residual{m}{ \Vector{\epsilon}_{#1} }

\NewDocumentCommand\SmallDelta{}{\scaleobj{0.8}{\Delta}}

\NewDocumentCommand\StateVector{}{ \Vector{\mathcal{X}} }
\NewDocumentCommand\StatePerturbation{}{ \delta\StateVector{} }

\NewDocumentCommand\SmallStateVector{}{\scaleobj{0.85}{\StateVector}}

\definecolor{rstablue}{cmyk}{1,0.1,0,0.1}
\definecolor{rstared}{cmyk}{0,1,0.8,0.1}

\tikzset{
  pointed/.style={
    -{Stealth[length=8pt,width=5pt]},
    line cap=round
  }
}

\newcommand{\drawFibre}[5][]{%
  \filldraw[
    gray!20!white,
    fill opacity=#5,
    draw=black,
    line width=0.5pt
  ]
    (tpp cs:x= 0,y= 0,z=#4) --
    (tpp cs:x= 0,y=#3,z=#4) --
    (tpp cs:x=#2,y=#3,z=#4) --
    (tpp cs:x=#2,y= 0,z=#4) -- cycle;

  \if\relax\detokenize{#1}\relax
  \else
    \node[anchor=west] at (tpp cs:x=#2 + 0.3,y=0,z=#4) {#1};
  \fi
}

\newcommand{\drawAxis}[4][]{%
  \foreach \xA/\yA/\zA in {#2}{
    \foreach \xB/\yB/\zB in {#3}{
      \draw[->,line width=0.75pt,black]
        (tpp cs:x=\xA,y=\yA,z=\zA) --
        (tpp cs:x=\xB,y=\yB,z=\zB);

      \if\relax\detokenize{#1}\relax
      \else
        \node[below]
          at (tpp cs:x=\xA - 0.1,y=\yA,z=\zA) {#1};
      \fi

      \foreach \zoff in {#4} {
        \draw[line width=0.75pt,black]
          (tpp cs:x=\xA - 0.15,y=\yA,z=\zoff) --
          (tpp cs:x=\xA + 0.15,y=\yA,z=\zoff);
      }
    }
  }
}

\newcommand{\drawStraightSegment}[5][]{%
  \foreach \xA/\yA/\zA in {#2}{}%
  \foreach \xB/\yB/\zB in {#3}{}%

  \pgfmathsetmacro{\ta}{(#4 - \zA)/(\zB - \zA)}%
  \pgfmathsetmacro{\tb}{(#5 - \zA)/(\zB - \zA)}%

  \pgfmathsetmacro{\xsa}{(1 - \ta)*\xA + \ta*\xB}%
  \pgfmathsetmacro{\ysa}{(1 - \ta)*\yA + \ta*\yB}%
  \pgfmathsetmacro{\zsa}{(1 - \ta)*\zA + \ta*\zB}%

  \pgfmathsetmacro{\xsb}{(1 - \tb)*\xA + \tb*\xB}%
  \pgfmathsetmacro{\ysb}{(1 - \tb)*\yA + \tb*\yB}%
  \pgfmathsetmacro{\zsb}{(1 - \tb)*\zA + \tb*\zB}%

  \if\relax\detokenize{#1}\relax
    \draw[thick]
      (tpp cs:x=\xsa,y=\ysa,z=\zsa) --
      (tpp cs:x=\xsb,y=\ysb,z=\zsb);
  \else
    \draw[#1]
      (tpp cs:x=\xsa,y=\ysa,z=\zsa) --
      (tpp cs:x=\xsb,y=\ysb,z=\zsb);
  \fi
}

\newcommand{\drawCurvedSegment}[7][]{%
  \pgfmathsetmacro{\Nsamp}{64}
  \pgfmathsetmacro{\dt}{1/(\Nsamp)}

  \pgfmathsetmacro{\xp}{\pgfmathresult}
  \foreach \xA/\yA/\zA in {#2}{}
  \foreach \xC/\yC/\zC in {#3}{}
  \foreach \xB/\yB/\zB in {#4}{}

  \pgfmathsetmacro{\xp}{\xA}
  \pgfmathsetmacro{\yp}{\yA}
  \pgfmathsetmacro{\zp}{\zA}

  \foreach \k in {1,...,\Nsamp} {
    \pgfmathsetmacro{\t}{\k*\dt}

    \pgfmathsetmacro{\xq}{
      (1 - \t)^2*\xA + 2*(1 - \t)*\t*\xC + \t^2*\xB}
    \pgfmathsetmacro{\yq}{
      (1 - \t)^2*\yA + 2*(1 - \t)*\t*\yC + \t^2*\yB}
    \pgfmathsetmacro{\zq}{
      (1 - \t)^2*\zA + 2*(1 - \t)*\t*\zC + \t^2*\zB}

    \ifdim \zp pt < #6 pt
      \ifdim \zq pt > #5 pt
        \ifnum\k = \Nsamp
          \if\relax\detokenize{#7}\relax
            \draw[thick,#1]
          \else
            \draw[thick,#1,#7]
          \fi
        \else
          \draw[thick,#1]
        \fi      
          (tpp cs:x=\xp,y=\yp,z=\zp) --
          (tpp cs:x=\xq,y=\yq,z=\zq);
      \fi
    \fi

    \xdef\xp{\xq}
    \xdef\yp{\yq}
    \xdef\zp{\zq}
  }
}

\newcommand{\drawStraightLineAtZ}[6][]{%
  \foreach \xA/\yA/\zA in {#2}{}%
  \foreach \xB/\yB/\zB in {#3}{}%
  \foreach \xC/\yC/\zC in {#4}{}%
  \foreach \xD/\yD/\zD in {#5}{}%

  \pgfmathsetmacro{\ta}{(#6 - \zA)/(\zB - \zA)}%
  \pgfmathsetmacro{\tb}{(#6 - \zC)/(\zD - \zC)}%

  \pgfmathsetmacro{\xa}{(1 - \ta)*\xA + \ta*\xB}%
  \pgfmathsetmacro{\ya}{(1 - \ta)*\yA + \ta*\yB}%

  \pgfmathsetmacro{\xb}{(1 - \tb)*\xC + \tb*\xD}%
  \pgfmathsetmacro{\yb}{(1 - \tb)*\yC + \tb*\yD}%

  \if\relax\detokenize{#1}\relax
    \draw[dashed,black]
      (tpp cs:x=\xa,y=\ya,z=#6) --
      (tpp cs:x=\xb,y=\yb,z=#6);
  \else
    \draw[#1]
      (tpp cs:x=\xa,y=\ya,z=#6) --
      (tpp cs:x=\xb,y=\yb,z=#6);
  \fi
}

\newcommand{\drawStraightEvent}[4][]{%
  \foreach \xA/\yA/\zA in {#2}{}
  \foreach \xB/\yB/\zB in {#3}{}

  \pgfmathsetmacro{\tb}{(#4 - \zA)/(\zB - \zA)}

  \pgfmathsetmacro{\xsb}{(1 - \tb)*\xA + \tb*\xB}
  \pgfmathsetmacro{\ysb}{(1 - \tb)*\yA + \tb*\yB}

  \if\relax\detokenize{#1}\relax
    \fill[black]
      (tpp cs:x=\xsb,y=\ysb,z=#4) circle (1.2pt);
  \else
    \fill[#1]
      (tpp cs:x=\xsb,y=\ysb,z=#4) circle (1.2pt);
  \fi
}

\newcommand{\drawCurvedEvent}[5][]{%
  \pgfmathsetmacro{\Nsamp}{64}
  \pgfmathsetmacro{\dt}{1/(\Nsamp)}

  \pgfmathsetmacro{\xp}{\pgfmathresult}
  \foreach \xA/\yA/\zA in {#2}{}
  \foreach \xC/\yC/\zC in {#3}{}
  \foreach \xB/\yB/\zB in {#4}{}

  \pgfmathsetmacro{\xp}{\xA}
  \pgfmathsetmacro{\yp}{\yA}
  \pgfmathsetmacro{\zp}{\zA}

  \pgfmathsetmacro{\eps}{0.025}

  \foreach \k in {1,...,\Nsamp} {
    \pgfmathsetmacro{\t}{\k*\dt}

    \pgfmathsetmacro{\xq}{
      (1 - \t)^2*\xA + 2*(1 - \t)*\t*\xC + \t^2*\xB}
    \pgfmathsetmacro{\yq}{
      (1 - \t)^2*\yA + 2*(1 - \t)*\t*\yC + \t^2*\yB}
    \pgfmathsetmacro{\zq}{
      (1 - \t)^2*\zA + 2*(1 - \t)*\t*\zC + \t^2*\zB}

    \pgfmathparse{abs(\zq-#5) <= \eps}
    \ifnum\pgfmathresult=1
      \if\relax\detokenize{#1}\relax
        \fill[black]
          (tpp cs:x=\xq,y=\yq,z=\zq) circle (1.2pt);
      \else
        \fill[#1]
          (tpp cs:x=\xq,y=\yq,z=\zq) circle (1.2pt);
      \fi
    \fi

    \xdef\xp{\xq}
    \xdef\yp{\yq}
    \xdef\zp{\zq}
  }
}

\newcommand{\drawStraightFrame}[7][]{%
  \foreach \xA/\yA/\zA in {#2}{}
  \foreach \xB/\yB/\zB in {#3}{}

  \pgfmathsetmacro{\tb}{(#4 - \zA)/(\zB - \zA)}
  \pgfmathsetmacro{\xsb}{(1 - \tb)*\xA + \tb*\xB}
  \pgfmathsetmacro{\ysb}{(1 - \tb)*\yA + \tb*\yB}

  \pgfmathsetmacro{\c}{cos(#5)}  %
  \pgfmathsetmacro{\s}{sin(#5)}

  \pgfmathsetmacro{\xr}{\xsb + #6*\c}
  \pgfmathsetmacro{\yr}{\ysb + #6*\s}

  \pgfmathsetmacro{\xg}{\xsb - #6*\s}
  \pgfmathsetmacro{\yg}{\ysb + #6*\c}

  \draw[->, line width=0.8pt, red,
      >={Stealth[length=5pt, width=4.0pt]}]
    (tpp cs:x=\xsb,y=\ysb,z=#4) -- (tpp cs:x=\xr,y=\yr,z=#4);

  \draw[->, line width=0.8pt, green!60!black,
      >={Stealth[length=5pt, width=3.0pt]}]
    (tpp cs:x=\xsb,y=\ysb,z=#4) -- (tpp cs:x=\xg,y=\yg,z=#4);
}

\titlehead{Research}

\begin{document}

\title{Uncertainty in space, time, and motion on the special Galilean group}

\author{Jonathan Kelly$^{1}$ and Matthew Giamou$^{2}$}

\address{%
$^{1}$Institute for Aerospace Studies, University of Toronto, Toronto, Canada\\
$^{2}$Department of Computing and Software, McMaster University, Hamilton, Canada}

\subject{Lie groups, mechanics}


\corres{Jonathan Kelly\\
\email{js.kelly@utoronto.ca}}

\AtBeginEnvironment{abstract}{
  \emergencystretch=2em
}

\begin{abstract}
Classical mechanics unfolds within absolute time and Euclidean space, yet our knowledge of \emph{where} events occur, \emph{when} they occur, and \emph{how} motion evolves is inherently uncertain.
The special Galilean group provides a natural setting for describing classical spacetime, combining absolute time, Euclidean space, and inertial motion within a single Lie group structure.
Although this framework is well known, representing and propagating uncertainty on the group has received comparatively little attention.
In this work, we bring together existing results on the structure of the Galilean group and use this unified framework to express uncertainty directly on the group manifold.
A main contribution is a compact, closed-form expression for the Galilean group Jacobian, which enables principled uncertainty propagation when composing Galilean transformations.
We show that uncertainty in spatial position and orientation, temporal displacement, and inertial motion are intrinsically coupled through the underlying group structure.
To illustrate the usefulness of the Galilean framework, we consider the problem of estimating a time-varying transformation between inertial frames from noisy observations collected at distinct instants in time.
We show that performing estimation directly on the Galilean group yields substantially more statistically consistent estimates than formulations that treat time independently.
Together, these results provide a geometric foundation for reasoning about uncertainty in space, time, and motion in classical mechanics, navigation, and robotics.
\end{abstract}

\begin{fmtext}

\end{fmtext}
\maketitle

\section{Introduction}

Inertial reference frames provide the basic geometric setting for classical mechanics.
An inertial frame is one in which Newton's first law holds, and any frame moving with constant velocity relative to an inertial frame is also inertial.
Motion plays out against the backdrop of Galilean spacetime, a four-dimensional affine manifold that brings space and time together within a single geometric framework \cite{2005_Penrose_Road}.
Observations made by different inertial observers are related by Galilean transformations, which include spacetime translations, rotations of spatial coordinates, and Galilean velocity boosts \cite{2011_Holm_Geometric_Part_II}.
These transformations preserve spatial separations and absolute time intervals between events, defined as points in Galilean spacetime.
The set of all Galilean transformations forms the \emph{special Galilean group}, which is the symmetry group of Galilean relativity \cite{1971_Levy-Leblond_Galilei}.
This group is a ten-dimensional Lie group, denoted $\LieGroupSGal{3}$, that captures the mathematical structure underlying inertial motion.

Galilean relativity provides a useful model for the evolution of physical systems, but in its usual form it assumes exact knowledge of position, orientation, velocity, and time.
In practice, all such measurements are inherently uncertain, and these uncertainties are coupled: uncertainty in time propagates into uncertainty in position through velocity, and uncertainty in velocity depends on the accuracy of both spatial and temporal measurements.
The special Galilean group provides a geometric framework in which these coupled uncertainties can be represented in a principled way.
Geometric representations of uncertainty on Lie groups have been applied successfully in several areas, including rigid-body motion, navigation, and state estimation on the special Euclidean group $\LieGroupSE{3}$, for example \cite{2014_Barfoot_Associating,2024_Barfoot_State}.
However, $\LieGroupSGal{3}$ has received almost no explicit attention from this perspective, despite its central role in relating space, time, and motion.
Existing approaches typically model uncertainty in position, orientation, and velocity only (e.g., \cite{2011_Kelly_Visual,2022_Brossard_Associating,2023_Barrau_Geometry}), without accounting for how temporal uncertainty interacts with spatial estimates through the underlying Lie group structure.

In this paper, we examine the special Galilean group $\LieGroupSGal{3}$ in the context of uncertainty in space, time, and inertial motion.
Our first contribution is to assemble a detailed description of $\LieGroupSGal{3}$, collecting material that is distributed across the literature.
We develop the matrix representation, present the Lie algebra, and define the exponential and logarithmic maps and the adjoint representation.
We also give a closed-form expression for the Jacobian of the exponential map. 
This Jacobian has only recently appeared in work by one of the authors and by others \cite{2023_Kelly_Galilean,2025_Delama_Equivariant,2025_Barfoot_Integral}, but the expression presented here is more compact than those previously published. 
The Jacobian is required for linearization and for the propagation of uncertainty on the group.

Our second and primary contribution is a geometric treatment of uncertainty on $\LieGroupSGal{3}$. 
Using the tools developed above, we show how uncertainty in position, orientation, velocity, and time can be represented and propagated directly on the group.
To the best of our knowledge, a thorough examination of uncertainty on
$\LieGroupSGal{3}$ has not appeared previously.
We demonstrate the utility of this geometric approach through an example problem in which a time-varying transformation between inertial frames must be inferred from noisy observations.
We adopt a total least squares, or errors-in-variables, formulation on the group, motivated by the fact that there is no independent variable in this setting. 
This formulation may be of independent interest for uncertainty modelling on Lie groups when all variables are subject to noise.

The remainder of the paper is organized as follows.
In \Cref{sec:galilean_relativity}, we review Galilean spacetime, inertial reference frames, and the Galilean transformations that relate them.
In \Cref{sec:SGal_group}, we introduce $\LieGroupSGal{3}$, including its matrix representation, Lie algebra, exponential and logarithmic maps, adjoint representation, and the associated Jacobian.
In \Cref{sec:uncertainty}, we develop a geometric treatment of uncertainty on $\LieGroupSGal{3}$.
The resulting formulation is illustrated in \Cref{sec:estimation} for the time-varying Galilean transformation estimation problem.
Throughout, we restrict our attention to inertial motion, which is already sufficiently complex under uncertainty and serves to demonstrate the value of the approach; a broader treatment of Galilean kinematics and related topics can be found in \cite{2025_Mahony_Galilean}.

\section{Galilean relativity}
\label{sec:galilean_relativity}

The principle of Galilean relativity states that the laws of physics are the same for all observers in uniform motion, although they may assign different spatial and temporal coordinates to their observations. %
Before introducing the Lie group structure used later in the paper, we  summarize the geometric framework on which this principle rests: Galilean spacetime, inertial reference frames, and the Galilean transformation that relates one inertial frame to another.

\subsection{The structure of Galilean spacetime}
\label{subsec:spacetime}

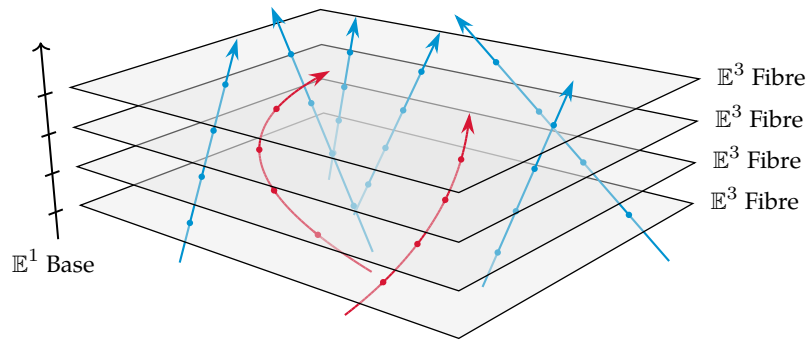
\begin{figure}[t!]
\centering
\pgfmathsetmacro{\elev}{ 27}
\pgfmathsetmacro{\azim}{-52}

\pgfmathsetmacro{\xext}{6.0}
\pgfmathsetmacro{\yext}{8.0}

\begin{tikzpicture}[
   3d view={\azim}{\elev},
   perspective={p={(30,0,0)}, q={(0,30,0)}, r={(0,0,-65)}}
]
  \def\straightlines{
    0.3/ 5.9/-0.6/ 1.9/6.0/2.8,
    4.5/ 6.5/-0.6/ 4.0/5.3/2.8,
    5.0/-0.2/-0.6/ 5.0/4.0/2.8,
    2.5/ 1.5/-0.6/ 3.0/0.3/2.8,
    3.5/ 5.0/-0.6/ 4.0/3.5/2.8,
    2.5/ 3.7/-0.6/ 3.7/7.0/2.8
  }

  \def\curvedlines{
     0.5/2.5/-0.6/ 1.0/0.6/2.8/ 1.2/0.9/1.0,
     1.9/3.2/-0.6/ 1.4/3.5/2.8/ 0.4/6.0/1.3
  }

  \def\zsteps{
    -0.6/0.0, %
     0.0/0.7,
     0.7/1.4,
     1.4/2.1,
     2.1/2.8  %
  }
  
  \foreach \za/\zb in \zsteps {
    \ifdim \zb pt < 2.8pt
      \foreach \xA/\yA/\zA/\xB/\yB/\zB in \straightlines {
         \drawStraightSegment[thick,rstablue]
         {\xA/\yA/\zA}{\xB/\yB/\zB}{\za}{\zb};
      }   

      \foreach \xA/\yA/\zA/\xB/\yB/\zB/\xC/\yC/\zC in \curvedlines {
        \drawCurvedSegment[thick,rstared]
        {\xA/\yA/\zA}{\xC/\yC/\zC}{\xB/\yB/\zB}{\za}{\zb}{}; %
      }
 
      \drawFibre[$\Euclid^{3}$ Fibre]{\xext}{\yext}{\zb}{0.4};
 
      \foreach \xA/\yA/\zA/\xB/\yB/\zB in \straightlines {
        \drawStraightEvent[rstablue]
        {\xA/\yA/\zA}{\xB/\yB/\zB}{\zb};
      }
      
      \foreach \xA/\yA/\zA/\xB/\yB/\zB/\xC/\yC/\zC in \curvedlines {
        \drawCurvedEvent[rstared]
        {\xA/\yA/\zA}{\xC/\yC/\zC}{\xB/\yB/\zB}{\zb};
      }
    \else
      \foreach \xA/\yA/\zA/\xB/\yB/\zB in \straightlines {
        \drawStraightSegment[thick,rstablue,pointed]
        {\xA/\yA/\zA}{\xB/\yB/\zB}{\za}{\zb};
      }   

      \foreach \xA/\yA/\zA/\xB/\yB/\zB/\xC/\yC/\zC in \curvedlines {
        \drawCurvedSegment[thick,rstared]
        {\xA/\yA/\zA}{\xC/\yC/\zC}{\xB/\yB/\zB}{\za}{\zb}{pointed};
      }
    \fi
  }

  \drawAxis[$\Euclid^{1}$ Base]
  {-0.3/8.3/-0.6}{-0.3/8.3/2.9}{-0.1,0.6,1.3,2.0};
\end{tikzpicture}
\caption{Galilean spacetime has the structure of a trivial fibre bundle with base space $\Euclid^{1}$ and fibres $\Euclid^{3}$.
The worldline of a body is a cross-section of the bundle, composed of a sequence of spacetime events (shown as dots).
A worldline is straight when the motion is inertial (blue), corresponding to a geodesic in Galilean spacetime.
Curved worldlines (red) indicate accelerated motion.
Figure inspired by \cite[Chapter 17]{2005_Penrose_Road}.}
\label{fig:spacetime}
\end{figure}

Galilean relativity unifies space and time into the four-dimensional affine manifold known as Galilean spacetime, $\Euclid^{1} \times \Euclid^{3}$, where $\Euclid^{1} \cong \Real$ denotes absolute time and $\Euclid^{3} \cong \Real^{3}$ denotes Euclidean space \cite{1981_Artz_Classical}.
A point in Galilean spacetime, called an event, lies in the the affine product $\Euclid^{1} \times \Euclid^{3}$, and may be expressed in coordinates as a tuple $(t, \Vector{x}) \in \Real \times \Real^{3}$.\footnote{An event and its coordinates are not the same thing, but we will treat them as synonymous.}
\Cref{fig:spacetime} illustrates the basic structure of Galilean spacetime as $\Euclid^{1} \times \Euclid^{3}$.

Galilean spacetime may be viewed as a (trivial) fibre bundle whose base space is absolute time $\Euclid^{1}$, where each instant $t$ is associated with a fibre $\Euclid^{3}$ representing all spatial points at that moment.
In this view, spacetime is a one-parameter family of Euclidean spaces indexed by time.
Spatial and temporal intervals are separate: all spatial displacements lie within a single $\Euclid^{3}$ fibre, while time differences belong to the base space $\Euclid^{1}$ \cite{2005_Penrose_Road}.
Moreover, Galilean spacetime has no preferred origin---that is, no privileged event that serves as a canonical `zero'.
The difference between two events (also called a translation) is determined by subtraction; this quantity is independent of the choice of coordinates or the existence of an origin.
However, events cannot be `added' in a meaningful way~\cite{2005_Bhand_Rigid}.

\subsection{Inertial reference frames and inertial motion}
\label{subsec:inertial_frames}

An inertial reference frame may be thought of as a standard Cartesian spatial frame (i.e., an orthogonal triad of coordinate axes) together with an origin from which spatial position and time are measured.
There is no privileged inertial frame, and hence all motion is relative.
However, as noted previously, time is absolute in Galilean relativity---all observers agree on a single, uniformly flowing time.
Moreover, simultaneity is absolute: if two events are simultaneous in one inertial frame, they remain simultaneous in all inertial frames 
\cite{1989_Arnold_Mathematical,2005_Bhand_Rigid}.
We denote an inertial reference frame by $\CoordinateFrame{a}$, where the
subscript identifies the frame.

The path of a body through Galilean spacetime is called its worldline.
This is a curve whose parameter is absolute time and whose values lie in the successive spatial fibres associated with each moment.
A worldline is straight precisely when the motion is inertial.
Formally, straight worldlines are the geodesics of the natural affine connection, which is flat and torsion-free.
Curvature in the worldline indicates that the body is accelerating (cf.\ \Cref{fig:spacetime}). 

\subsection{Galilean transformations}
\label{subsec:galilean_transformations}

Inertial reference frames are related by \emph{Galilean transformations},  composed of spatial rotations, spacetime translations, and Galilean boosts.
These operations span the full set of admissible coordinate changes between inertial frames. 
We briefly consider how each operation acts on an event before presenting the complete Galilean group structure in \Cref{sec:SGal_group}.

\subsubsection{Spatial rotations}
\label{subsubsec:rotations}

A spatial rotation is a linear transformation of spatial coordinates defined by an element of the \emph{special orthogonal group}.
The \emph{special orthogonal group} $\LieGroupSO{3}$ of rigid-body rotations is
\begin{equation}
\label{eqn:SO3_definition}
\LieGroupSO{3} \Defined
\left\{
\Matrix{C} \in \Real^{3 \times 3} \,\Big|\,
\Matrix{C}\Matrix{C}^{\T} = \Identity_{3}, \Determinant{\Matrix{C}} = 1
\right\},
\end{equation}
where $\Identity_{3}$ is the 3 $\times$ 3 identity matrix.
A rotation acts only on the spatial coordinates $\Vector{x}$ of an event $\left(t, \Vector{x}\right)$ via
\begin{equation}
(t, \Vector{x}) \mapsto (t, \Matrix{C}\Vector{x}).
\end{equation}
Because $\Matrix{C}$ is orthonormal, Euclidean distances are preserved.\footnote{Since $\Determinant{\Matrix{C}} = +1$, we consider proper rotations only, which preserve spatial handedness.}

Later, we will require the Lie algebra of $\LieGroupSO{3}$, denoted
$\LieAlgebraSO{3}$, whose elements have the form
\begin{equation}
\label{eqn:so3_definition}
\LieAlgebraSO{3} \Defined
\left\{
\Matrix{\Phi} = \Vector{\phi}^{\wedge} \in \Real^{3 \times 3}
\,\Big|\, \Vector{\phi} \in \Real^{3}
\right\}.
\end{equation}
The linear `wedge' operator $\Wedge{(\cdot)}: \Real^{3} \to \Real^{3 \times 3}$ is
\begin{equation}
\Vector{\phi}^{\wedge} =
\bbm
\phi_1 \\
\phi_2 \\
\phi_3
\ebm^{\wedge} = 
\bbm
0 & -\phi_3 & \phi_2 \\
\phi_3 & 0 & -\phi_1 \\
-\phi_2 & \phi_1 & 0
\ebm 
\in \mathbb{R}^{3 \times 3},\;
\Vector{\phi} \in \mathbb{R}^3,
\end{equation}
which produces a skew-symmetric matrix.
The `vee' operator $\Vee{(\cdot)}$ recovers $\Vector{\phi}$ from
$\Matrix{\Phi} = \Vector{\phi}^{\wedge}$,
\begin{equation*}
\Matrix{\Phi} = \Vector{\phi}^{\wedge}
\quad \longleftrightarrow \quad 
\Vector{\phi} = \Matrix{\Phi}^{\vee}.
\end{equation*}
A derivation of \Cref{eqn:so3_definition} appears in \cite[Chapter~4]{2005_Selig_Geometric} and elsewhere.

\subsubsection{Spacetime translations}
\label{subsubsec:translations}

The coordinates of an event $\left(t, \Vector{x}\right)$ may be translated in space and time by the pair $\left(\tau, \Vector{r}\right) \in \Real \times \Real^{3}$,
\begin{equation}
\left(t, \Vector{x}\right) \mapsto 
\left(t + \tau, \Vector{x} + \Vector{r}\right).
\end{equation}
Spatial translations are \emph{isochronous}, in the sense that shifting an event in space does not alter its time coordinate; similarly, time translations shift all events uniformly along the absolute time axis.
 
\subsubsection{Galilean boosts}
\label{subsubsec:boosts}

Inertial reference frames may undergo constant, rectilinear motion with respect to one another.
A \emph{Galilean boost} describes this relationship.
For a velocity $\Vector{v} \in \Real^{3}$, the action of a boost on an event $(t, \Vector{x})$ is
\begin{equation}
(t, \Vector{x}) \mapsto (t, \Vector{x} + \Vector{v}t).
\end{equation}
Only relative velocities matter---just as Galilean spacetime has no preferred origin, it also has no preferred state of motion or rest  \cite{2012_Maudlin_Philosophy}.

\section{The special Galilean Lie group}
\label{sec:SGal_group}

In this section we gather the Lie-theoretic tools needed to work with the special Galilean group. We introduce its matrix representation, describe the associated Lie algebra, and review the exponential and logarithmic maps, the adjoint representation, and the Jacobian of the group exponential.
We conclude by discussing how the group acts on itself and on spacetime events, together with the resulting geometric invariants.\footnote{Throughout, we work with the connected component at the identity of the group $\LieGroupGal{3}$, denoted $\LieGroupSGal{3}$.}

\subsection{The matrix representation of $\LieGroupSGal{3}$}

Elements of $\LieGroupSGal{3}$ can be written as 5$\times$5 matrices,
\begin{equation}
\label{eqn:SGal3_definition}
\LieGroupSGal{3} \Defined
\begin{Bmatrix}
\Matrix{F} =
\bbm 
\Matrix{C} & \Vector{v} & \Vector{r} \\ 
\Zero & 1 & \tau \\
\Zero & 0 & 1
\ebm
\in \Real^{5 \times 5}
\ \vast\vert \ 
\Matrix{C} \in \LieGroupSO{3}, 
\Vector{v} \in \Real^3,
\Vector{r} \in \Real^3,
\tau \in \Real
\end{Bmatrix},
\end{equation}
where $\Zero$ is a submatrix of zeros; when the size is clear from context, we omit any subscripts on the matrix $\Zero$.
We use $\Matrix{F} \in \LieGroupSGal{3}$ to denote an element of the Galilean group.
The inverse of $\Matrix{F}$ is
\begin{equation}
\label{eqn:SGal3_inverse}
\Inv{\Matrix{F}} =
\bbm
 \Transpose{\Matrix{C}} &
-\Transpose{\Matrix{C}}\Vector{v} &
-\Transpose{\Matrix{C}}\left(\Vector{r} - \Vector{v}\tau\right) \\ 
 \Zero & 1 & -\tau \\
 \Zero & 0 & 1
\ebm,
\end{equation}
such that $\Matrix{F}\Inv{\Matrix{F}}\! = \Identity_{5}$.
This matrix form defines an inclusion $\LieGroupSGal{3} \hookrightarrow \LieGroupGL{5,\Real}$, the group of invertible 5$\times$5 real matrices; the group operation is matrix multiplication.
The Galilean group can be decomposed as $\smash{\LieGroupSGal{3} \cong \big(\LieGroupSO{3} \ltimes \Real^{3}\big)} \ltimes \smash{\big(\Real \times \Real^{3} \big)}$, where the inner semidirect product corresponds to rotations and boosts, and the outer semidirect product corresponds to spacetime translations.

\subsection{The Lie algebra $\LieAlgebraSGal{3}$}
\label{subsec:sgal3_lie_algebra}

The set of all tangent vectors at the identity element of $\LieGroupSGal{3}$ defines its Lie algebra $\LieAlgebraSGal{3}$.
This tangent space is a 10-dimensional real vector space (i.e., equal to the dimension of the group).
Each element of the algebra is the tangent of a smooth curve on $\LieGroupSGal{3}$ through the identity, generating an infinitesimal Galilean transformation.

Elements of $\LieAlgebraSGal{3}$ can be written in block-matrix form as
\begin{equation}
\label{eqn:sgal3_definition}
\LieAlgebraSGal{3} \Defined
\begin{Bmatrix}
\Matrix{\Xi} = \hspace{1pt}
\bbm
\Vector{\phi}^{\wedge} & \Vector{\nu} & \Vector{\rho} \\
\Zero & 0 & \iota \\
\Zero & 0 & 0
\ebm
\in \Real^{5 \times 5}
\ \vast\vert \
\Vector{\phi}  \in \Real^3,
\Vector{\nu}   \in \Real^3,
\Vector{\rho}  \in \Real^3,
\iota \in \Real
\end{Bmatrix}.
\end{equation}
For convenience, we overload the $\Wedge{(\cdot)}$ operator in the usual way \cite{1994_Murray_Mathematical},
\begin{equation}
\label{eqn:sgal_wedge}
\Vector{\xi}^{\wedge}
\Defined
\bbm
\Vector{\rho} \\
\Vector{\nu}  \\
\Vector{\phi} \\
\iota
\ebm^{\wedge}=
\bbm
\Vector{\phi}^{\wedge} & \Vector{\nu} & \Vector{\rho} \\
\Zero & 0 & \iota \\
\Zero & 0 & 0
\ebm \in \LieAlgebraSGal{3},
\end{equation}
as a mapping $\Real^{10} \rightarrow \LieAlgebraSGal{3}$.
Similarly, we overload the corresponding $\Vee{(\cdot)}$ operator such that
\begin{equation*}
\Vector{\xi}^{\wedge} = \Matrix{\Xi}
\quad \longleftrightarrow \quad
\Matrix{\Xi}^{\vee} = \Vector{\xi}.
\end{equation*}

Since $\LieAlgebraSGal{3}$ is 10-dimensional, its elements can be expressed as linear combinations of ten basis elements (generators), corresponding respectively to rotations, boosts, spatial translations, and time translations.
Their explicit forms follow directly from \eqref{eqn:sgal_wedge} by setting each parameter to unity in turn.
A full derivation of the Lie algebra $\LieAlgebraSGal{3}$, starting from the group structure, is provided in \cite{2023_Kelly_Galilean}.

The Lie bracket of two elements $\Matrix{\Xi}_{1}, \Matrix{\Xi}_{2} \in \LieAlgebraSGal{3}$ quantifies the degree of noncommutativity of infinitesimal transformations and is given by the matrix commutator
\begin{equation}
[\Matrix{\Xi}_{1}, \Matrix{\Xi}_{2}]
=
\Matrix{\Xi}_{1}\,\Matrix{\Xi}_{2} - 
\Matrix{\Xi}_{2}\,\Matrix{\Xi}_{1} \in \LieAlgebraSGal{3}.
\end{equation}
This operation is closed within the algebra.
See \cite{2011_Holm_Geometric_Part_II, 1999_Marsden_Introduction} for further details, and \cite{2005_Selig_Geometric} for additional discussion of the Lie bracket.

\subsection{The exponential and logarithmic maps}
\label{subsec:exp_log_maps}

We next describe how to move between $\LieAlgebraSGal{3}$ and $\LieGroupSGal{3}$.
The exponential and logarithmic maps define a local diffeomorphism between the tangent space at the identity element and the group manifold \cite{2011_Chirikjian_Stochastic}.
These maps enable linearization in the algebra and integration on the group.
The exponential map from $\LieAlgebraSGal{3}$ to $\LieGroupSGal{3}$ is
\begin{equation}
\label{eqn:SGal3_exp_map}
\exp\left(\Vector{\xi}^{\wedge}\right)
=
\sum_{n = 0}^{\infty}\frac{1}{n!}
\bbm
\Vector{\phi}^{\wedge} & \Vector{\nu} & \Vector{\rho} \\
\Zero & 0 & \iota \\
\Zero & 0 & 0
\ebm^{n} =
\bbm
\Matrix{C}\left(\Vector{\phi}\right) & 
\Matrix{D}\left(\Vector{\phi}\right)\<\Vector{\nu} & 
\Matrix{D}\left(\Vector{\phi}\right)\Vector{\rho} +
\Matrix{E}\left(\Vector{\phi}\right)\<\Vector{\nu}\iota \\
\Zero & 1 & \iota \\
\Zero & 0 & 1
\ebm,
\end{equation}
where the matrices $\Matrix{C}$, $\Matrix{D}$, and $\Matrix{E}$ are given in closed form below.

To evaluate the exponential, we first consider the rotation submatrix $\Matrix{C}$, obtained from the exponential map on $\LieGroupSO{3}$.
Using the axis-angle parameterization $\Vector{\phi} = \phi\<\Vector{u}$, where $\phi = \Norm{\Vector{\phi}}$ is the angle of rotation about the unit-length axis $\Vector{u} = \Vector{\phi}/\Norm{\Vector{\phi}}$, we have
\begin{align}
\label{eqn:SO3_exp_map}
\Matrix{C}\left(\Vector{\phi}\right)
= 
\exp\left(\phi\<\uw\right)
= 
\sum_{n = 0}^{\infty}
\frac{1}{n!}\big(\phi\<\uw\big)^n
=
\Identity_{3} +
\sin\left(\phi\right)\uw +
\bigl(1 - \cos\left(\phi\right)\bigr)\uw\uw,
\end{align}
which is the well-known Rodrigues rotation formula \cite[Chapter 2.2]{1994_Murray_Mathematical}.
The map from $\LieAlgebraSO{3}$ to $\LieGroupSO{3}$ is surjective but not injective: adding any integer multiple of $2\pi$ to $\phi$ yields the same $\Matrix{C}$.
The remaining matrices $\Matrix{D}$ and $\Matrix{E}$ admit the closed-form expressions
\begin{align}
\label{eqn:D_exp_map}
\Matrix{D}\left(\Vector{\phi}\right)
=
\sum_{n = 0}^{\infty}
\frac{1}{(n + 1)!}\big(\phi\<\uw\big)^n
=
\Identity_{3} + 
\left(\frac{1 - \cos\left(\phi\right)}{\phi}\right)\uw +
\left(\frac{\phi - \sin\left(\phi\right)}{\phi}\right)\uw\uw
\end{align}
and
\begin{align}
\label{eqn:E_exp_map}
\Matrix{E}\left(\Vector{\phi}\right)
=
\sum_{n = 0}^{\infty}
\frac{1}{(n + 2)!}\big(\phi\<\uw\big)^n
=
\frac{1}{2}\Identity_{3} +
\left(\frac{\phi - \sin\left(\phi\right)}{\phi^{2}}\right)\uw +
\left(\frac{\phi^{2} + 2\cos\left(\phi\right) - 2}{2\phi^{2}}\right)\uw\uw.
\end{align}

Deriving the logarithmic map from $\LieGroupSGal{3}$ to $\LieAlgebraSGal{3}$ requires some additional steps.
First, the rotation angle $\phi$ is recovered from the matrix trace,
\begin{equation}
\label{eqn:angle_from_C}
\phi =
\cos^{-1}\left(\frac{\Trace{\Matrix{C}} - 1}{2}\right),
\end{equation}
which is likewise not unique; uniqueness can be enforced by choosing $\phi$ such that $\Norm{\Vector{\phi}} < \pi$.
The logarithmic map from $\LieGroupSO{3}$ to $\LieAlgebraSO{3}$ is 
\begin{equation}
\label{eqn:axis_angle_from_C}
\Vector{\phi} =
\Matlog{\Matrix{C}}^{\vee} =
\left(\frac{\phi}{2\sin\left(\phi\right)}
\left(\Matrix{C} - \Matrix{C}^{\T}\right)\right)^{\vee},
\end{equation}
with $\Vector{u}^{\wedge} = \ln\left(\Matrix{C}\right)/\phi$.
We also require the inverse of $\Matrix{D}$, which has the closed-form expression
\begin{equation}
\label{eqn:D_inverse_closed_form}
\Inv{\Matrix{D}}\left(\Vector{\phi}\right)
=
\Identity_{3} -
\frac{\phi}{2}\Matrix{u}^{\wedge} +
\Bigg(1 - \frac{\phi}{2}\cot\left(\frac{\phi}{2}\right)\Bigg)
\Vector{u}^{\wedge}\Vector{u}^{\wedge}.
\end{equation}
Deriving the full logarithmic map yields expressions for the translational and boost components in terms of the recovered rotation parameters and the matrices $\Matrix{D}$ and $\Matrix{E}$.
The result is
\begin{equation}
\label{eqn:SGal3_log_map}
\Vector{\xi}
=
\Matlog{\Matrix{F}}^{\vee}
=
\bbm
\Inv{\Matrix{D}}\left(\Vector{r} - \Matrix{E}\,\Vector{\nu}\iota\right) \\[0.1mm]
\Inv{\Matrix{D}}\Vector{v} \\[0.1mm]
\Matlog{\Matrix{C}}^{\vee} \\
\tau
\ebm.
\end{equation}

Notably, because Lie group multiplication is non-commutative in general, a product of exponentials cannot be written as the exponential of a simple sum in the Lie algebra.
This behaviour is captured by the Baker-Campbell-Hausdorff (BCH) formula, which expresses the product of two exponentials as a single exponential whose argument is an infinite series of nested Lie brackets.
For $\Vector{\xi},\< \Vector{\zeta} \in \LieAlgebraSGal{3}$ with sufficiently small norm, the BCH formula gives
\begin{equation}
\label{eqn:bch_formula}
\ln\!\Big(\!
\Matexp{\Vector{\xi}^{\wedge}}
\Matexp{\Vector{\zeta}^{\wedge}}
\Big)^{\vee}
=
\Vector{\xi} +
\Vector{\zeta} +
\frac{1}{2}[\Vector{\xi},\Vector{\zeta}] +
\frac{1}{12}
\left([\Vector{\xi},[\Vector{\xi},\Vector{\zeta}]] +
[\Vector{\zeta},[\Vector{\zeta},\Vector{\xi}]]\right)
+ \dots,
\end{equation}
where $[\cdot, \cdot]$ denotes the Lie bracket on $\LieAlgebraSGal{3}$ and the ellipsis indicates higher-order nested commutator terms.
This expansion makes explicit how non-commutativity of the group gives rise to nested Lie bracket terms in the Lie algebra.

The BCH formula links group composition and algebraic addition, and provides the theoretical foundation for using the exponential and logarithmic maps in local computations on the group.
In \Cref{sec:estimation}, we use truncated forms of this expansion in our estimation framework to obtain linear approximations for small perturbations.

\subsection{The adjoint map and the adjoint representation}
\label{subsec:adjoint}

It is often necessary to map a Lie algebra element (a tangent-space vector) from the tangent space at one group element to the tangent space at another.
For Lie groups, this transformation is linear. 
The linear action of a group on its Lie algebra is called the \emph{adjoint representation} of the group.
In the case of $\LieGroupSGal{3}$, this is the map $\mathrm{Ad}_{\Matrix{F}} : \LieAlgebraSGal{3} \rightarrow \LieAlgebraSGal{3}$, acting on tangent-space vectors.
To derive this map, we follow \cite[Section II.F]{2021_Sola_Micro}, starting from the identity 
\begin{equation}
\label{eqn:adjoint_identity}
\begin{aligned}
\Matexp{\left(\LieGroupAdjoint{\Matrix{F}}\Vector{\xi}\right)^{\wedge}}
\Matrix{F}
& =
\Matrix{F}\exp\left(\Vector{\xi}^{\wedge}\right) \\[1mm]
\Matexp{\left(\LieGroupAdjoint{\Matrix{F}}\Vector{\xi}\right)^{\wedge}}
& =
\Matrix{F}\exp\left(\Vector{\xi}^{\wedge}\right)\Inv{\Matrix{F}} \\
\LieGroupAdjoint{\Matrix{F}}{\Vector{\xi}}
& =
\left(\Matrix{F}\Vector{\xi}^{\wedge}\Matrix{F}^{-1}\right)^{\vee}.
\end{aligned}
\end{equation}
The expression above for the adjoint defines a mapping from the tangent space at $\Matrix{F}$ (the local frame, on the right) to the tangent  space at the identity (the global frame, on the left).

Because the adjoint action is linear in $\Vector{\xi}$, it may be expressed as a 10$\times$10 matrix with respect to the chosen basis of $\LieAlgebraSGal{3}$.
The adjoint matrix is given by
\begin{equation}
\label{eqn:SGal3_adjoint_definition}
\LieGroupAdjoint{\Matrix{F}}
\Defined
\bbm
 \Matrix{C}\left(\Vector{\phi}\right) &
-\Matrix{C}\left(\Vector{\phi}\right)\tau &
 \left(\Vector{r} - \Vector{v}\tau\right)^{\wedge}
 \Matrix{C}\left(\Vector{\phi}\right) &
 \Vector{v} \\
 \Zero & \Matrix{C}\left(\Vector{\phi}\right) &
 \Vector{v}^{\wedge}\Matrix{C}\left(\Vector{\phi}\right) &
 \Zero \\
 \Zero &
 \Zero &
 \Matrix{C}\left(\Vector{\phi}\right) & 
 \Zero \\
 \Zero &
 \Zero &
 \Zero & 1
\ebm
\in \Real^{10 \times 10}.
\end{equation}

Analogously to the group case, the Lie algebra $\LieAlgebraSGal{3}$ admits a representation on itself via the Lie bracket.
This representation is the linear map $\mathrm{ad}_{\Matrix{\Xi}} : \LieAlgebraSGal{3} \rightarrow \LieAlgebraSGal{3}$.\footnote{The lowercase $\mathrm{ad}$ notation distinguishes the Lie algebra adjoint from the Lie group adjoint, $\mathrm{Ad}$.}
To determine the form of this map, we begin with the Lie bracket,
\begin{equation}
\label{eqn:sgal3_adjoint_derivation}
\LieAlgebraAdjoint{\Matrix{\Xi_{1}}}\Matrix{\Xi}_{2}
=
[\Matrix{\Xi}_{1}, \Matrix{\Xi}_{2}] 
=
\Matrix{\Xi}_{1}\Matrix{\Xi}_{2} - \Matrix{\Xi}_{2}\Matrix{\Xi}_{1}.
\end{equation}
The corresponding adjoint matrix for $\Matrix{\Xi} = \Vector{\xi}^{\wedge}$ is
\begin{equation}
\label{eqn:sgal3_adjoint_definition}
\LieAlgebraAdjoint{\Vector{\xi}^{\wedge}}
\Defined
\bbm
 \Vector{\phi}^{\wedge} &
-\Identity_{3}\iota     & 
 \Vector{\rho}^{\wedge} & 
 \Vector{\nu} \\
\Zero & \Vector{\phi}^{\wedge} & \Vector{\nu}^{\wedge} & \Zero \\
\Zero & \Zero & \Vector{\phi}^{\wedge} & \Zero \\
\Zero & \Zero & \Zero & 0
\ebm
\in \Real^{10 \times 10}.
\end{equation}
The group and algebra adjoints are related by
\begin{equation}
\LieGroupAdjoint{\exp\left(\Vector{\xi}^{\wedge}\right) %
\rule{0pt}{10pt}}          %
=
\exp\left(\LieAlgebraAdjoint{\Vector{\xi}^{\wedge}} %
\rule{0pt}{10pt}\right),  %
\end{equation}
which makes explicit how the adjoint actions are connected via the exponential map.

\subsection{The Jacobian}
\label{subsec:jacobian}

To linearize perturbations on $\LieGroupSGal{3}$ and to propagate uncertainty through compositions of transformations, we require the Jacobian of the exponential map.
The Jacobian relates small perturbations in the Lie algebra to first-order perturbations on the group.
Let $\Vector{\xi}^{\wedge} \in \LieAlgebraSGal{3}$ denote an element of the Lie algebra and let $\delta\Vector{\xi}^{\wedge}$ denote a small perturbation.
Using a first-order truncation of the BCH formula to linearize this perturbation on the group, we obtain
\begin{equation}
\label{eqn:bch_left_jacobian}
\Matexp{\left(\Vector{\xi} + \delta\Vector{\xi}\right)^{\wedge}}
\approx
\Matexp{
\left(
\LeftJacobianSGal\left(\Vector{\xi}\right)
\delta\Vector{\xi}
\right)^{\wedge}
}
\Matexp{\Vector{\xi}^{\wedge}},
\end{equation}
where $\LeftJacobianSGal\left(\Vector{\xi}\right)$ is the \emph{left} Jacobian, a linear map $\Matrix{J}_\ell(\boldsymbol{\xi}):\LieAlgebraSGal{3} \rightarrow \LieAlgebraSGal{3}$.
Explicitly, the left Jacobian can be shown to be
\begin{equation}
\label{eqn:SGal3_left_jacobian}
\LeftJacobianSGal\left(\Vector{\xi}\right)
=
\int_{0}^{1}
\exp\left(
\Vector{\xi}^{\wedge}
\right)^\alpha d\alpha
=
\sum_{n = 0}^{\infty} 
\frac{1}{(n + 1)!}
\LieAlgebraAdjoint{\Vector{\xi}^{\wedge}}^{n}.
\end{equation}
Alternatively, we can apply the perturbation `on the right',
\begin{equation}
\label{eqn:bch_right_jacobian}
\Matexp{\left(\Vector{\xi} + \delta\Vector{\xi}\right)^{\wedge}}
\approx
\Matexp{\Vector{\xi}^{\wedge}}
\Matexp{
\left(
\RightJacobianSGal\left(\Vector{\xi}\right)
\delta\Vector{\xi}
\right)^{\wedge}
},
\end{equation}
where $\RightJacobianSGal\left(\Vector{\xi}\right)$ is the \emph{right} Jacobian. The two Jacobians are related by $\RightJacobianSGal\left(\Vector{\xi}\right) = \LeftJacobianSGal\left(-\Vector{\xi}\right)$.

The left Jacobian has the closed-form expression
\begin{equation}
\label{eqn:SGal3_left_jacobian_definition}
\Matrix{J}_{\ell}\left(\Vector{\xi}\right)
\Defined
\bbm
 \Matrix{D}\left(\Vector{\phi}\right) &
-\Matrix{L}\left(\Vector{\phi}\right)\<\iota & 
 \Matrix{N}\left(\Vector{\phi}, \Vector{\nu}, \Vector{\rho}\right) & 
 \Matrix{E}\left(\Vector{\phi}\right)\<\Vector{\nu} \\
\Zero &
\Matrix{D}\left(\Vector{\phi}\right) &
\Matrix{M}\left(\Vector{\phi}, \Vector{\nu}\right) &
\Zero \\
\Zero & 
\Zero &
\Matrix{D}\left(\Vector{\phi}\right) &
\Zero \\
\Zero &
\Zero &
\Zero &
1
\ebm
\in \Real^{10 \times 10}.
\end{equation}
The matrices $\Matrix{D}$ and $\Matrix{E}$ are given by \Cref{eqn:D_exp_map} and  \Cref{eqn:E_exp_map}, respectively. The matrix $\Matrix{L}$ is
\begin{equation}
\label{eqn:L_jacobian_short}
\begin{aligned}
\Matrix{L}\left(\Vector{\phi}\right)
& =
\sum_{n = 0}^{\infty}
\frac{n + 1}{(n + 2)!}\big(\phi\<\uw\big)^n \\
& = 
\frac{1}{2}\Identity_{3} +
\left( 
\frac{\sin\left(\phi\right) - \phi\cos\left(\phi\right)}
{\phi^{2}}
\right)
\uw +
\left(
\frac{\phi^{2} + 
 2 - 2\phi\sin\left(\phi\right) - 2\cos\left(\phi\right)}
{2\<\phi^{2}}
\right)
\uw\uw.
\end{aligned}
\end{equation}
The matrix $\Matrix{M}$ is
\begin{equation}
\begin{aligned}
\Matrix{M}\left(\Vector{\phi}, \Vector{\nu}\right)
& =
\sum_{n = 0}^{\infty}
\sum_{m = 0}^{\infty}
\frac{1}{(n + m + 2)!}\phi^{n + m}
\left(\uw\right)^{n}
\nuw
\left(\uw\right)^{m} \\
& =
\left(\frac{1 - \cos\left(\phi\right)}{\phi^2}\right)
\nuw +
\left(
\frac{\phi - \sin\left(\phi\right)}{\phi^2}
\right)
\left(
\uw
\nuw +
\nuw
\uw
\right) \\
& +
\left(
\frac{2 - \phi\sin\left(\phi\right) - 2\cos\left(\phi\right)}{\phi^2}
\right)
\uw
\nuw
\uw +
\left(
\frac{2\phi + \phi\cos\left(\phi\right) - 3\sin\left(\phi\right)}{\phi^2}
\right)
\uw
\nuw 
\uw
\uw.
\end{aligned}	
\end{equation}
Lastly, the matrix $\Matrix{N}$ is most easily expressed as the difference of two individual matrices, as
\begin{equation}
\Matrix{N}\left(\Vector{\phi}, \Vector{\nu}, \Vector{\rho}\right) 
=
\Matrix{N}_{1}\left(\Vector{\phi}, \Vector{\rho}\right) - 
\Matrix{N}_{2}\left(\Vector{\phi}, \Vector{\nu}\right).
\end{equation}
The matrix $\Matrix{N}_{1}$ is
\begin{align}
\label{eqn:N1_jacobian_short}
\Matrix{N}_{1}\left(\Vector{\phi}, \Vector{\rho}\right)
& =
\sum_{n = 0}^{\infty}
\sum_{m = 0}^{\infty}
\frac{1}{(n + m + 2)!}\phi^{n + m}
\left(\uw\right)^{n}
\puw
\left(\uw\right)^{m} \notag \\
&
\begin{multlined}[b]
=
\left(\frac{1 - \cos\left(\phi\right)}{\phi^2}\right)
\puw
+
\left(
\frac{\phi - \sin\left(\phi\right)}{\phi^2}
\right)
\left(
\uw
\puw +
\puw
\uw
\right) \\
+
\left(
\frac{2 - \phi\sin\left(\phi\right) - 2\cos\left(\phi\right)}{\phi^2}
\right)
\uw
\puw
\uw +
\left(
\frac{2\phi + \phi\cos\left(\phi\right) - 3\sin\left(\phi\right)}{\phi^2}
\right)
\uw
\puw 
\uw
\uw,
\end{multlined}
\end{align}
which appears as part of the Jacobian of the group $\LieGroupSE{3}$.
The matrix $\Matrix{N}_{2}$ is
\begin{align}
\Matrix{N}_{2}\left(\Vector{\phi}, \Vector{\nu}\right)
& = 
\sum_{n = 0}^{\infty}
\sum_{m = 0}^{\infty}
\frac{n + 1}{(n + m + 3)!}\phi^{n + m}
\left(\uw\right)^{n}
\nuw\iota
\left(\uw\right)^{m} \notag \\
&
\begin{multlined}[b]
=
\Bigg(
\left(
\frac{2 - \phi\sin\left(\phi\right) - 
2\cos\left(\phi\right)}{\phi^{3}}
\right)
\uw +
\left(
\frac{\phi + \phi\cos\left(\phi\right) - 
2\sin\left(\phi\right)}{\phi^{3}}
\right)
\uw
\uw
\Bigg)
\nuw\iota \hspace{0.5in} \\ %
+
\left(
\frac{
4\sin\left(\phi\right) -
 \phi^{2}\sin\left(\phi\right) -
4\phi\cos\left(\phi\right)}
{2\<\phi^3}
\right)
\uw
\nuw
\uw
\iota \hspace{0.94in} \\ %
\hspace{0.71in}
+
\left(
\frac{
4 +
\phi^{2} +
\phi^{2}\cos\left(\phi\right) -
4\phi\sin\left(\phi\right) -
4\cos\left(\phi\right)
}
{2\<\phi^{3}}
\right)
\uw
\nuw
\uw
\uw
\iota \\
+
\nuw\iota
\Bigg(
\left(
\frac{\phi^{2} + 2\cos\left(\phi\right) - 2}{2\<\phi^{3}}
\right)
\uw +
\left(
\frac{\sin\left(\phi\right) - \phi}{\phi^{3}}
\right)
\uw
\uw
\Bigg).
\end{multlined}
\end{align}
To the best of our knowledge, this Jacobian first appeared in the technical report \cite{2023_Kelly_Galilean} by one of the present authors, and the form given here is more compact than others found in the literature.

\subsection{The group action and geometric invariants}

We consider two actions of the Galilean group: its action on itself through composition, and its action on spacetime events.
When working with multiple inertial reference frames, we adopt the convention that $\Matrix{F}_{ab} \in \LieGroupSGal{3}$ denotes the transformation that maps coordinates expressed in frame $\CoordinateFrame{b}$ into coordinates expressed in frame $\CoordinateFrame{a}$.\footnote{The order of the subscripts indicates the direction of mapping and should be read from right to left.}
We use this convention consistently throughout the paper.

Consider three inertial reference frames, $\CoordinateFrame{a}$, $\CoordinateFrame{b}$, and $\CoordinateFrame{g}$. 
A compound transformation from frame $\CoordinateFrame{b}$ to
$\CoordinateFrame{g}$ is obtained by left-multiplication:
\begin{equation}
\label{eqn:sgal3_composition}
\Matrix{F}_{gb} 
=
\Matrix{F}_{ga}\Matrix{F}_{ab},
\quad
\Matrix{F}_{gb},
\Matrix{F}_{ga},
\Matrix{F}_{ab} \in \LieGroupSGal{3}.
\end{equation}
This composition rule follows the familiar case of rigid-body
$\LieGroupSE{3}$ transformations. 
In \Cref{sec:estimation}, we consider transformations of the form in \Cref{eqn:sgal3_composition} but with uncertainty.

To describe the action of the group on events, we model Galilean spacetime as the set
\begin{equation}
\label{eqn:event_definition}
\mathcal{E} \Defined
\left\{
\Vector{p} =
\begin{bmatrix}
\Vector{x}^{\T}\nds & t\! & 1
\end{bmatrix}^{\T}
\;\Bigg|\;
\Vector{x} \in \Real^{3},\ t \in \Real
\right\},
\end{equation}
where each $\Vector{p} \in \mathcal{E}$ is a homogeneous coordinate representation of an event $(t,\Vector{x})$ in spacetime.
The set $\mathcal{E}$ thus contains all events expressed in homogeneous coordinates.
For $\Matrix{F}_{ab} \in \LieGroupSGal{3}$ of the form~\eqref{eqn:SGal3_definition}, the action of the group on the homogeneous coordinates of an event is given by matrix multiplication,
\begin{equation}
\label{eqn:event_action}
\Vector{p}_{a}
=
\Matrix{F}_{ab}\<\Vector{p}_{b}
=
\begin{bmatrix}
\Matrix{C} & \Vector{v} & \Vector{r} \\
\Zero & 1 & \tau \\
\Zero & 0 & 1
\end{bmatrix}
\begin{bmatrix}
\Vector{x} \\[2pt] t \\[2pt] 1
\end{bmatrix}
=
\begin{bmatrix}
\Matrix{C}\Vector{x} + \Vector{v}\,t + \Vector{r} \\
t + \tau \\
1
\end{bmatrix},
\end{equation}
which maps the coordinates of an event from frame $\CoordinateFrame{b}$ to frame $\CoordinateFrame{a}$.\footnote{The Galilean group also acts on directions and velocities, in addition to events. Further details are available in \cite{2025_Mahony_Galilean}.}

Galilean transformations define the affine symmetries of Galilean spacetime.
They preserve the underlying affine structure while changing spatial and temporal coordinates, and leave unchanged a small set of geometric quantities:
\begin{itemize}
\setlength{\itemsep}{2pt}
\item the time interval $\lvert t_{b} - t_{a} \rvert$ between any two events $(t_{a}, \Vector{x}_{a})$ and $(t_{b}, \Vector{x}_{b})$,
\item the Euclidean distance $\|\Vector{x}_{b} - \Vector{x}_{a}\|_{2}$ between events at the same time (critically), and
\item the handedness of space, because only proper rotations are permitted.
\end{itemize}
Although Galilean inertial reference frames and Galilean transformations share the same algebraic structure, they play different roles: Galilean transformations are elements of $\LieGroupSGal{3}$, whereas Galilean reference frames are elements of the group torsor.\footnote{An $\LieGroupSGal{3}$-torsor is a set whose elements differ by Galilean transformations, but without a preferred origin or identity.}

\section{Uncertainty on $\LieGroupSGal{3}$}
\label{sec:uncertainty}

We can express the uncertainty associated with an element of $\LieGroupSGal{3}$ in terms of a perturbation in the tangent space at the identity.
To do so, we follow the approach in \cite{2014_Barfoot_Associating} and assume that the perturbation is a zero-mean vector-valued Gaussian random variable, $\Vector{\xi} \sim \NormalDistribution{\Vector{0}}{\Matrix{\Sigma}}$, $\Vector{\xi} \in \Real^{10}$, where $\Matrix{\Sigma} \in \Real^{10 \times 10}$ is symmetric positive definite.
The perturbation can be applied locally (on the right) or globally (on the left),
\begin{equation}
\Matrix{F} = \Mean{\Matrix{F}}\exp\left(\Vector{\xi}^{\wedge}\right)
\quad\text{or}\quad
\Matrix{F} = \exp\left(\Vector{\xi}^{\wedge}\right)\Mean{\Matrix{F}},
\end{equation}
respectively, where $\Mean{\Matrix{F}}$ is a `large' deterministic transformation.
For $\LieGroupSGal{3}$, the probability density function (PDF) of the perturbation is
\begin{equation}
p(\Vector{\xi})
=
\eta
\exp\left(
-\frac{1}{2}\Vector{\xi}^{\T}\Matrix{\Sigma}^{-1}\Vector{\xi}
\right),
\quad
\eta = \frac{1}{\sqrt{(2\pi)^{10}\det(\Matrix{\Sigma})}}.
\end{equation}

The distribution $p(\Vector{\xi})$ in the tangent space induces a corresponding distribution $p(\Matrix{F})$ on the group via the exponential map.
To relate these distributions, we note that the right Jacobian of the exponential map on $\LieGroupSGal{3}$,
$\RightJacobianSGal(\Vector{\xi})$, determines the change in volume between the tangent space and the group.
In particular, the induced infinitesimal volume element on $\LieGroupSGal{3}$ satisfies
\begin{equation}
d\<\Matrix{F}
=
\left|
\Determinant{\RightJacobianSGal\left(\Vector{\xi}\right)}
\right|\,d\Vector{\xi},
\end{equation}
Since $\LieGroupSGal{3}$ is unimodular, the determinants of the left and right Jacobians coincide, so either may be used to express this volume scaling. We adopt the right Jacobian because our estimation framework in \Cref{sec:estimation} is expressed in right-trivialized coordinates.

To determine the density on the group, we apply a change of variables. Under our perturbation model,
\begin{equation}
\Matrix{F}
=
\Mean{\Matrix{F}}\exp\!\left(\Vector{\xi}^{\wedge}\right),
\end{equation}
we have locally
\begin{equation}
\Matlog{\Mean{\Matrix{F}}^{-1}\Matrix{F}}^{\vee}
=
\Vector{\xi},
\end{equation}
and hence
\begin{equation}
\Expectation{\Vector{\xi}}
=
\Expectation{
\Matlog{\Mean{\Matrix{F}}^{-1}\Matrix{F}}^{\vee}
}
=
\Vector{0},
\end{equation}
so $\Mean{\Matrix{F}}$ is the (right-invariant) mean of the distribution.
Conservation of probability mass implies
\begin{equation}
p(\Matrix{F})\< d\Matrix{F}
=
p(\Vector{\xi})\< d\Vector{\xi},
\end{equation}
and substituting the volume relation gives the induced density on the group, leading to
\begin{align}
1
& =
\int_{\Real^{10}}
\eta\,
\exp\left(
-\frac{1}{2}
\Vector{\xi}^{\T}\Matrix{\Sigma}^{-1}\Vector{\xi}
\right)
d\Vector{\xi} \\
& =
\int_{\LieGroupSGal{3}}
\beta\,
\exp\left(
-\frac{1}{2}
\Transpose{\Matlog{\Inv{\Mean{\Matrix{F}}}\Matrix{F}}^{\vee}}
\Matrix{\Sigma}^{-1}
\Matlog{\Inv{\Mean{\Matrix{F}}}\Matrix{F}}^{\vee}
\right)
\,d\Matrix{F} \\
& =
\int_{\LieGroupSGal{3}} p(\Matrix{F})\,d\Matrix{F},
\end{align}
where $\beta$ is the normalization factor arising from the change of variables,
\begin{equation}
\label{eq:sgal3_beta}
\beta
=
\frac{\eta}{
\left|\det\left(\RightJacobianSGal(\Vector{\xi})\right)\right|
}
=
\frac{1}{
\sqrt{(2\pi)^{10}
\Determinant{
\RightJacobianSGal(\Vector{\xi})\,
\Matrix{\Sigma}\,
\RightJacobianSGal(\Vector{\xi})^{\T}
}}}.
\end{equation}
This follows the approach of Barfoot and Furgale~\cite{2014_Barfoot_Associating}, but it is not the only way to define a mean and covariance on a Lie group; in particular, other choices can yield a normalization factor that is constant instead of depending on $\Matrix{F}$~\cite{2011_Wolfe_Bayesian}.

\begin{figure}[t!]
\centering
\includegraphics[width=\textwidth]{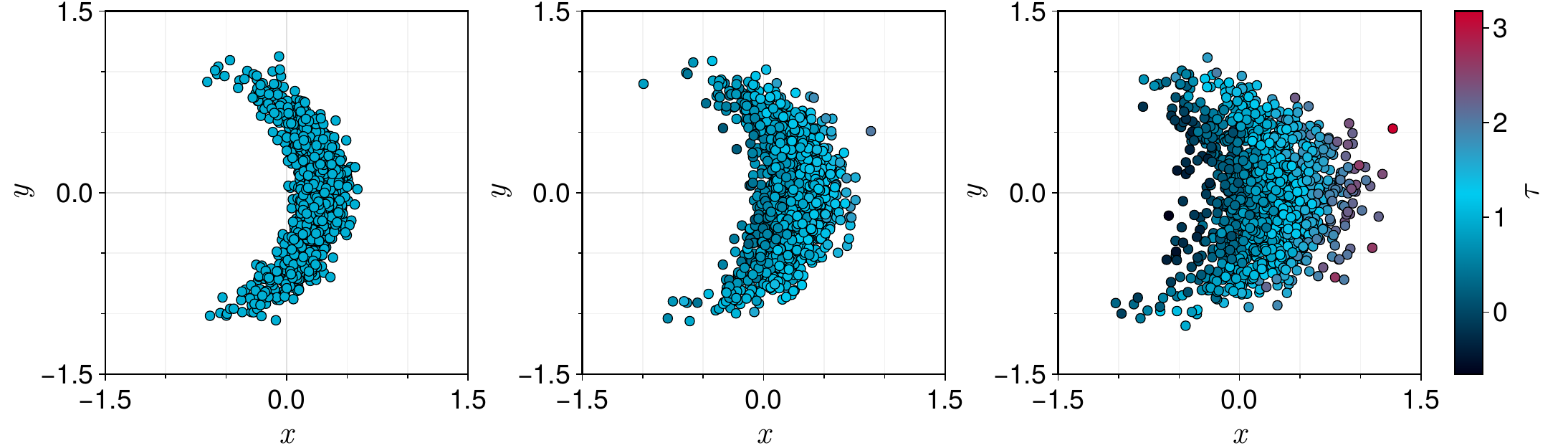}
\caption{Visualization of the transformation of an event by a right-perturbed element of $\LieGroupSGal{3}$, projected onto the $x$-$y$ plane. Left: perturbation to $x$ translation and $z$ rotation components only. Middle: additional (small) perturbation in time. Right: additional (large) perturbation in time. Points are shaded by time offset, according to the colour bar shown on the right. Temporal uncertainty induces a `spread' in the spatial uncertainty. Each plot shows 1,200 samples drawn from a multivariate Gaussian.}
\label{fig:uncertainty}
\vspace{-0.25\baselineskip}
\end{figure}

\Cref{fig:uncertainty} illustrates how the introduction of temporal uncertainty modifies the familiar `banana-shaped' spatial uncertainty that arises from rotational and translational noise alone \cite{2012_Long_Banana}.
In each panel, 1,200 randomly sampled events specified in the local frame are mapped to the global frame by a Galilean transformation whose mean specifies a translation in $x$ together with a nonzero velocity in the positive $x$ direction.
Uncertainty is introduced via a right-perturbation of this mean element of $\LieGroupSGal{3}$, and the resulting events are then projected onto the $x$-$y$-plane for visualization.
The perturbations in yaw and $x$-translation are modelled as uncorrelated zero-mean Gaussian noise in the tangent space, with standard deviations of $\pi$/6 and 0.1, respectively.

With perturbations to yaw and $x$-translation only (left panel), the nonlinear action of the group `bends' the Gaussian into the characteristic banana-shaped distribution in the plane, as is well known from the $\LieGroupSE{3}$ case.
When temporal uncertainty is added (centre and right panels, with temporal noise standard deviations of 0.3 and 0.6, respectively), the coupling between time, velocity, and spatial displacement induces an additional spread of the samples along the direction of motion.\footnote{By analogy with the banana distribution, we call the resulting distribution the `dumpling distribution'.}
We emphasize that the uncertainty in the tangent space \emph{is} Gaussian, but appears non-elliptic after exponentiation and projection onto the $x$--$y$ plane.

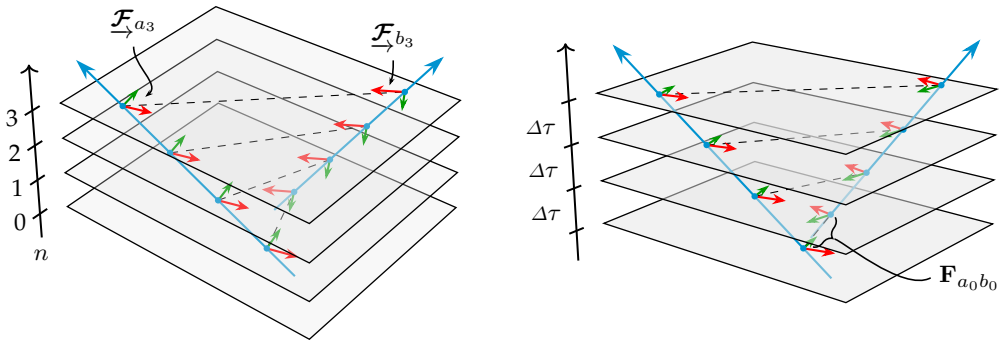
\begin{figure}[t!]
\centering
\begin{minipage}{0.48\textwidth}
  \centering
  \pgfmathsetmacro{\elev}{ 45}
\pgfmathsetmacro{\azim}{-52}

\pgfmathsetmacro{\xext}{3.5}
\pgfmathsetmacro{\yext}{4.67}

\begin{tikzpicture}[
   3d view={\azim}{\elev},
   perspective={p={(30,0,0)}, q={(0,30,0)}, r={(0,0,-65)}}
]
  \def\straightlines{
    1.2/1.2/-0.4/ 0.3/4.6/2.8 / -75,
    2.2/2.4/-0.4/ 3.5/0.3/2.8 / 120
  }

  \def\zsteps{
    -0.4/0.0, %
     0.0/0.7,
     0.7/1.4,
     1.4/2.1,
     2.1/2.8  %
  }
  
  \foreach \za/\zb in \zsteps {
    \ifdim \zb pt < 2.8pt
      \foreach \xA/\yA/\zA/\xB/\yB/\zB/\ang in \straightlines {
        \drawStraightSegment[thick,rstablue]
        {\xA/\yA/\zA}{\xB/\yB/\zB}{\za}{\zb};
      }   

      \drawFibre{\xext}{\yext}{\zb}{0.25};
      
      \foreach [count=\i] \xA/\yA/\zA/\xB/\yB/\zB/\ang in \straightlines {
        \foreach [count=\j] \xC/\yC/\zC/\xD/\yD/\zD/\ang in \straightlines {
          \ifnum\j>\i
          	\drawStraightLineAtZ[dashed,black]
          	{\xA/\yA/\zA}{\xB/\yB/\zB}{\xC/\yC/\zC}{\xD/\yD/\zD}{\zb};
          \fi
        }
      }

      \foreach \xA/\yA/\zA/\xB/\yB/\zB/\ang in \straightlines {
        \drawStraightFrame[opacity=0.9]
        {\xA/\yA/\zA}{\xB/\yB/\zB}{\zb}{\ang}{0.5}

        \drawStraightEvent[rstablue]
        {\xA/\yA/\zA}{\xB/\yB/\zB}{\zb};
      }
    \else
      \foreach \xA/\yA/\zA/\xB/\yB/\zB/\ang in \straightlines {
        \drawStraightSegment[thick,rstablue,pointed]
        {\xA/\yA/\zA}{\xB/\yB/\zB}{\za}{\zb};
      }
    \fi
  }

  \drawAxis[$\,\,n$]
  {-0.3/4.97/-0.6}{-0.3/4.97/2.9}{-0.1,0.6,1.3,2.0};
  
  \foreach \z/\lab in {-0.1/0,0.6/1,1.3/2,2.0/3} {
    \node[anchor=east,inner sep=1pt]
      at (tpp cs:x=-0.3-0.15-0.07,y=4.97,z=\z-0.01) {\lab};
  }

  \node[anchor=south] (OA) at (2.9,5.2) {$\CoordinateFrame{a_{3}}$};
  \coordinate (ptA) at (2.25,4.5);
  \draw[->,line width=0.6pt,>={Stealth[length=3pt,width=2pt]}]
  (OA.south) to[out=-90,in=40] (ptA);
  
  \node[anchor=south] (OB) at (4.8,2.35) {$\CoordinateFrame{b_{3}}$};
  \coordinate (ptB) at (4.4,2.1);
  \draw[->,line width=0.6pt,>={Stealth[length=3pt,width=2pt]}]
  (OB.south) to[out=-90,in=70] (ptB);
  
\end{tikzpicture}
\end{minipage}\hfill
\begin{minipage}{0.48\textwidth}
  \centering
  \pgfmathsetmacro{\elev}{ 25}
\pgfmathsetmacro{\azim}{-52}

\pgfmathsetmacro{\xext}{3.5}
\pgfmathsetmacro{\yext}{4.67}

\begin{tikzpicture}[
   3d view={\azim}{\elev},
   perspective={p={(30,0,0)}, q={(0,30,0)}, r={(0,0,-65)}}
]
  \def\straightlines{
    1.2/1.2/-0.4/ 0.3/4.6/2.8 / -75,
    2.2/2.4/-0.4/ 3.5/0.3/2.8 /  80
  }

  \def\zsteps{
    -0.4/0.0, %
     0.0/0.7,
     0.7/1.4,
     1.4/2.1,
     2.1/2.8  %
  }
  
  \foreach \za/\zb in \zsteps {
    \ifdim \zb pt < 2.8pt
      \foreach \xA/\yA/\zA/\xB/\yB/\zB/\ang in \straightlines {
        \drawStraightSegment[thick,rstablue]
        {\xA/\yA/\zA}{\xB/\yB/\zB}{\za}{\zb};
      }   

      \drawFibre{\xext}{\yext}{\zb}{0.5};
      
      \foreach [count=\i] \xA/\yA/\zA/\xB/\yB/\zB/\ang in \straightlines {
        \foreach [count=\j] \xC/\yC/\zC/\xD/\yD/\zD/\ang in \straightlines {
          \ifnum\j>\i
          	\drawStraightLineAtZ[dashed,black]
          	{\xA/\yA/\zA}{\xB/\yB/\zB}{\xC/\yC/\zC}{\xD/\yD/\zD}{\zb};
          \fi
        }
      }

      \foreach \xA/\yA/\zA/\xB/\yB/\zB/\ang in \straightlines {
        \drawStraightFrame[opacity=0.9]
        {\xA/\yA/\zA}{\xB/\yB/\zB}{\zb}{\ang}{0.5}

        \drawStraightEvent[rstablue]
        {\xA/\yA/\zA}{\xB/\yB/\zB}{\zb};
      }
    \else
      \foreach \xA/\yA/\zA/\xB/\yB/\zB/\ang in \straightlines {
        \drawStraightSegment[thick,rstablue,pointed]
        {\xA/\yA/\zA}{\xB/\yB/\zB}{\za}{\zb};
      }
    \fi
  }

  \drawAxis[]
  {-0.3/4.97/-0.6}{-0.3/4.97/2.9}{-0.1,0.6,1.3,2.0};

  \foreach \z/\lab in 
  {0.25/$\SmallDelta\tau$,0.95/$\SmallDelta\tau$,1.65/$\SmallDelta\tau$} {
    \node[anchor=east,inner sep=1pt]
      at (tpp cs:x=-0.3-0.15-0.07,y=4.97,z=\z) {\lab};
  }

  \begin{scope}[shift={(0,1)},rotate=25]
    \draw[decorate,decoration={brace,mirror,amplitude=4pt},line width=0.6pt]
    (0.85,0) -- (1.55,0) coordinate[pos=0.48] (ptBrace); 
  \end{scope}
  
  \node[anchor=west] (label) at (1.3,-0.4) {$\Matrix{F}_{a_{0}b_{0}}$};

  \draw[line width=0.6pt] (label.west) .. 
    controls +(-0.6,0) and +(0.08,0) .. ($(ptBrace)+(4.05pt,-2.45pt)$);
\end{tikzpicture}
\end{minipage}
\caption{Visualization of the problem considered in \Cref{sec:estimation}. Two spacetime diagrams show inertial frames $\smash{\CoordinateFrame{a}}$ and $\smash{\CoordinateFrame{b}}$ evolving along distinct worldlines, as viewed from a third inertial frame $\CoordinateFrame{g}$ fixed in space and time.
Because $\smash{\CoordinateFrame{a}}$ and $\smash{\CoordinateFrame{b}}$ are in relative motion, their spatial separation varies over time (grey dashed lines).
Left: observations occur at discrete time indices $n$, where paired measurements of position, orientation, velocity, and local time are obtained at $\smash{\CoordinateFrame{a_n}}\!$ and $\smash{\CoordinateFrame{b_n}}\!$; frames $\smash{\CoordinateFrame{a_3}}\!$ and $\smash{\CoordinateFrame{b_3}}\!$ are highlighted.
Right: an alternate perspective. At each time slice (fibre), the spatial frame axes are drawn in red ($x$) and green ($y$); these axes remain fixed in orientation under inertial motion.
The estimation task is to find the Galilean transformation $\Matrix{F}_{a_{0}b_{0}}$ relating $\smash{\CoordinateFrame{b_0}}$ to $\smash{\CoordinateFrame{a_0}}$, along with the inter-measurement time $\SmallDelta\tau$, using only the paired observations along each worldline.}
\label{fig:evolving_frames}
\vspace{-0.25\baselineskip}
\end{figure}

\section{Estimation on $\LieGroupSGal{3}$}
\label{sec:estimation}

In this section, we apply the geometric and probabilistic machinery developed above to a representative estimation problem. 
The example is chosen to illustrate how Galilean transformations arise naturally when determining relative motion and timing from noisy observations and to highlight the interplay between spatial and temporal uncertainty.
To the best of the authors' knowledge, with the exception of the recent treatment in \cite{2025_Mahony_Galilean}, earlier work on estimation involving Galilean transformations is limited to \cite{2021_Giefer_Uncertainties}, which considers $\LieGroupSGal{2}$ only and deals with a different estimation task.

We consider the problem of estimating the spatiotemporal relationship between two inertial reference frames, denoted $\CoordinateFrame{a}$ and $\CoordinateFrame{b}$, both of which evolve along inertial worldlines through time.
At discrete and unknown but regular sampling intervals, we obtain measurements of position, velocity, orientation, and time along each worldline, expressed relative to a third inertial reference frame $\CoordinateFrame{g}$ at a chosen time instant.
The inertial reference frames themselves are elements of the group torsor, while transformations between frames are elements of $\LieGroupSGal{3}$.
We illustrate the problem setup in \Cref{fig:evolving_frames}. 

This problem has several features that make it well suited for illustrating important aspects of Galilean transformations in estimation.
The group structure makes explicit the time dependence of spatial relationships between frames and allows these relationships to be propagated forward or backward in time in a principled way.
From an estimation-theoretic perspective, the problem naturally leads to a total least squares (or errors-in-variables) formulation on $\LieGroupSGal{3}$, since both spatial and temporal quantities are measured with uncertainty.
The group adjoint and associated Jacobians provide the appropriate tools for linearization and optimization on this space.
Although real trajectories may deviate from ideal inertial motion, this formulation is sufficient to demonstrate how the Galilean group framework supports the treatment of coupled spatial and temporal uncertainty.

\subsection{System modelling}
\label{subsec:system_modelling}

We describe the system model in terms of the following components: the initial transformation model, the motion model, the observation model, and the perturbation model used for estimation, introducing notation as needed.
Each component is described in detail in the sections that follow.

The fundamental deterministic relationship between the frames $\CoordinateFrame{a}$ and $\CoordinateFrame{b}$ at some initial time is given by
\begin{equation}
\Matrix{F}_{gb_{0}} 
=
\Matrix{F}_{ga_{0}}
\Matrix{F}_{a_{0}b_{0}}.
\end{equation}
We extend our transformation notation by appending an integer subscript to the frame identifiers to indicate the observation index.
The integer index $n = 0, \dots, N$ denotes the sequence of observations only, not the exact acquisition times. 
In particular, the index $n = 0$ denotes an unknown but fixed initial transformation relating the frames along their worldlines.
The index is ordered monotonically with time; for convenience, we take increasing $n$ to correspond to increasing time, though the formulation is agnostic to this choice.

We assume that the relative transformation $\Matrix{F}_{a_{0}b_{0}}$ is unknown, but that an initial estimate is available.
Likewise, the time interval between successive observations is unknown but fixed across all samples.
This reflects a common situation in tracking and estimation problems, where measurements are acquired at regular but (possibly) uncertain intervals.
All observations are assumed to be acquired simultaneously across frames at each index $n$.
The estimation problem is therefore to determine both the time increment $\SmallDelta\tau$ and the initial relative transform $\Matrix{F}_{a_{0}b_{0}}$ from noisy observations along the two worldlines.

\subsubsection{Motion model}
\label{subsubsec:motion_model}

We adopt an inertial motion model throughout this work.\footnote{While the
framework can be extended to more general motions, inertial trajectories are
sufficient for the present example.}
The reference frames $\CoordinateFrame{a}$ and $\CoordinateFrame{b}$ (or the observers attached to these frames) are assumed to move along straight worldlines.
When expressed in its own local coordinates, an inertial frame therefore evolves only through the passage of time, without any spatial displacement.

From a geometric perspective, moving forward or backward along a worldline
corresponds to composition with a pure time translation. 
For inertial motion, the temporal coordinate is the only local degree of freedom that changes between samples.
As a result, the Galilean transformation between successive samples along a single worldline consists of a time translation only.
Using the discrete sampling index $n$, we write
\begin{equation*}
\Matrix{F}_{a_{1}a_{0}}=\Matrix{F}_{\nhs\SmallDelta\tau},
\quad
\Matrix{F}_{b_{1}b_{0}}=\Matrix{F}_{\nhs\SmallDelta\tau},
\end{equation*}
and more generally $\Matrix{F}_{a_{n}a_{0}} = \Matrix{F}_{\nhs\SmallDelta\tau}^{n}$ and $\Matrix{F}_{b_{n}b_{0}} = \Matrix{F}_{\nhs\SmallDelta\tau}^{n}$, where the time-translation element in $\LieGroupSGal{3}$ has the form
\begin{equation}
\Matrix{F}_{\!\SmallDelta\tau}
=
\bbm
\Identity_{3} & \Zero & \Zero \\
\Zero & 1 & \SmallDelta\tau \\
\Zero & 0 & 1
\ebm.
\end{equation}

The relationship between the two frames $\CoordinateFrame{a}$ and $\CoordinateFrame{b}$, as viewed from the global frame $\CoordinateFrame{g}$, is then determined by
\begin{equation}
\label{eqn:motion_model}
\Matrix{F}_{gb_{n}}
\Matrix{F}_{\nhs\SmallDelta\tau}^{n}
=
\Matrix{F}_{ga_{n}}
\Matrix{F}_{\nhs\SmallDelta\tau}^{n}\,
\Matrix{F}_{a_{0}b_{0}},
\end{equation}
which can equivalently be written as
\begin{equation}
\Matrix{F}_{gb_{n}}
=
\Matrix{F}_{ga_{n}}
\left(
\Matrix{F}_{\nhs\SmallDelta\tau}^{n}\,
\Matrix{F}_{a_{0}b_{0}}\,
\Matrix{F}_{\nhs\SmallDelta\tau}^{-n}
\right).
\end{equation}
This shows that the relative transform between the frames at sample index $n$ is obtained by conjugating the initial transform $\Matrix{F}_{a_{0}b_{0}}$ by the time translation $\Matrix{F}_{\nhs\SmallDelta\tau}^{n}$. 
The conjugation reflects the fact that the spatial relationship between inertial frames depends on the time at which it is evaluated. In practice, the sign of $\SmallDelta\tau$ is determined by the temporal ordering of the samples.

\subsubsection{Observation model}
\label{subsubsec:observation_model}

This section defines the observation model linking noisy measurements to the underlying Galilean transformations.
We describe how measurements are represented on the group and how uncertainty is modelled in a form suitable for estimation.

For each sample index $n$, let $\Observation{\Matrix{F}}_{ga_n}$ and $\Observation{\Matrix{F}}_{gb_n}$ denote the measured transformations from the inertial frames $\CoordinateFrame{a}$ and $\CoordinateFrame{b}$, respectively, to the global frame $\CoordinateFrame{g}$.
We assume that all components of the Galilean transformation are observed, including orientation, position, velocity, and time offset.
This is the most informative observation model, but the formulation readily accommodates partial observations by omitting unmeasured components, without altering the basic structure of the problem.
In particular, a partial observation model could involve observing events only, with measurements restricted to position and time.

We model measurement noise using left-invariant residuals,
\begin{equation}
\label{eqn:left_residuals}
\Residual{a_{n}} =
\Matlog{\Inv{\Observation{\Matrix{F}}_{ga_{n}}}\Matrix{F}_{ga_{n}}}^{\vee}
\sim \NormalDistribution{\Vector{0}}{\Matrix{\Sigma}_{a_{n}}},
\quad
\Residual{b_{n}} =
\Matlog{\Inv{\Observation{\Matrix{F}}_{gb_{n}}}\Matrix{F}_{gb_{n}}}^{\vee}
\sim \NormalDistribution{\Vector{0}}{\Matrix{\Sigma}_{b_{n}}},
\end{equation}
which quantify the difference between the true and observed transformations in left-invariant coordinates.
Here, we use a tilde to distinguish observed (measured) quantities from their underlying true values.
We assume Gaussian measurement noise for analytical convenience; although
temporal errors are likely to be non-Gaussian in practice, this approximation is sufficient for our development here.

Substituting the residual definitions into the motion model
\Cref{eqn:motion_model} and rearranging terms yields
\begin{equation}
\Observation{\Matrix{F}}_{gb_{n}}
\Matexp{-\Residual{b_{n}}^{\wedge}}\,
\Matrix{F}_{\nhs\SmallDelta\tau}^{n}
=
\Observation{\Matrix{F}}_{ga_{n}}
\Matexp{-\Residual{a_{n}}^{\wedge}}\,
\Matrix{F}_{\nhs\SmallDelta\tau}^{n}\,
\Matrix{F}_{a_{0}b_{0}},
\end{equation}
which defines one constraint for each pair of observations.
This expression makes explicit how temporal uncertainty directly influences the inferred spatial relationship between frames: errors in time measurements propagate into spatial errors through inertial motion.

\subsubsection{Perturbation model}

To formulate our total least squares problem on $\LieGroupSGal{3}$, we introduce local perturbations in the tangent space of the group.
We first perturb the unknown relative transform $\Matrix{F}_{a_{0}b_{0}}$.
Let $\smash{\Prior{\Matrix{F}}_{a_{0}b_{0}}}\!$ denote the current linearization point; a $(\Prior{\cdot})$ is used throughout to indicate current estimates.
We write the perturbed transform in terms of a right-invariant Lie algebra perturbation $\delta\Vector{\xi}_{ab} \in \Real^{10}$ as
\begin{equation}
\label{eqn:base_perturbation}
\Matrix{F}_{a_{0}b_{0}}
=
\Prior{\Matrix{F}}_{a_{0}b_{0}}
\Matexp{\delta\Vector{\xi}_{ab}^{\wedge}},
\end{equation}
where an equivalent left-invariant formulation could also be used.

We next consider perturbations to the time-translation component.
Let $\Prior{\Matrix{F}}_{\nhs\SmallDelta\tau}$ denote the current linearization point for the time translation. We write the perturbed time translation in terms of a Lie algebra perturbation $\delta\Vector{\xi}_{\tau}$ as
\begin{equation}
\label{eqn:time_perturbation}
\Matrix{F}_{\SmallDelta\tau}
=
\Prior{\Matrix{F}}_{\nhs\SmallDelta\tau}
\Matexp{\delta\Vector{\xi}_{\tau}^{\wedge}},
\end{equation}
where
\begin{equation*}
\delta\Vector{\xi}_{\tau}
=
\bbm
\Zero_{1 \times 9} \nds & \delta\tau
\ebm^{\T}
\end{equation*}
and $\delta\tau$ is the scalar perturbation to $\SmallDelta\tau$.
Because the subgroup of spacetime translations is Abelian, time translations commute with one another and with their associated perturbations. As a result, the $n$-step time evolution under this perturbation satisfies
\begin{equation}
\Big(\Prior{\Matrix{F}}_{\nhs\SmallDelta\tau}
\Matexp{\delta\Vector{\xi}_{\tau}^{\wedge}}\Big)^{n} 
=
\Prior{\Matrix{F}}_{\nhs\SmallDelta\tau}^{n}
\Matexp{n\<\delta\Vector{\xi}_{\tau}^{\wedge}},
\end{equation}
where the equality follows directly from commutativity. This linear dependence of the propagated time perturbation on the sample index $n$ is a key structural property of the estimation problem, since a single unknown time increment influences all observations in a coupled way.

Substituting the perturbed transformations into the model yields
\begin{equation}
\label{eq:model_perturbed}
\Observation{\Matrix{F}}_{gb_{n}}\!
\Matexp{-\Residual{b_{n}}^{\wedge}}
\Prior{\Matrix{F}}_{\nhs\SmallDelta\tau}^{n}
\Matexp{n\<\delta\Vector{\xi}_{\tau}^{\wedge}}
=
\Observation{\Matrix{F}}_{ga_{n}}\!
\Matexp{-\Residual{a_{n}}^{\wedge}}
\Prior{\Matrix{F}}_{\nhs\SmallDelta\tau}^{n}
\Matexp{n\<\delta\Vector{\xi}_{\tau}^{\wedge}}
\Prior{\Matrix{F}}_{a_{0}b_{0}}
\Matexp{\delta\Vector{\xi}_{ab}^{\wedge}}.
\end{equation}
For each sample index $n$, \eqref{eq:model_perturbed} defines a set of nonlinear constraints in the perturbation variables.
The collection of these constraints over all samples forms the system to be linearized in the following section.
For convenience, we stack all the perturbation variables into a single vector
\begin{equation*}
\StatePerturbation
\Defined
\bbm
\delta\Vector{\xi}_{ab}^{\T} \nds & \delta\tau
\ebm^{\T}.
\end{equation*}

\subsection{Total least squares optimization}
\label{subsec:tls_optimization}

We now formulate the estimation problem as a total least squares (TLS) optimization on $\LieGroupSGal{3}$.
As discussed previously, in many estimation problems, the independent variable, typically time, is assumed to be noise-free.
In contrast, for estimation on the Galilean group, time itself is a measured quantity subject to uncertainty.
The problem therefore naturally takes an errors-in-variables form, in which spatial and temporal quantities are uncertain and must be estimated jointly.
We minimize the scalar cost function
\begin{equation}
\label{eqn:tls_cost}
S
=
\sum_{n = 0}^{N}
\frac{1}{2}
\left(
\Residual{a_{n}}^{\T}
\Matrix{\Sigma}_{a_{n}}^{-1}
\Residual{a_{n}}
+
\Residual{b_{n}}^{\T}
\Matrix{\Sigma}_{b_{n}}^{-1}
\Residual{b_{n}}
\right),
\end{equation}
where $\Residual{a_{n}}$ and $\Residual{b_{n}}$ are the left-invariant observation residuals defined by \Cref{eqn:left_residuals}, and $\Matrix{\Sigma}_{a_{n}}$ and $\Matrix{\Sigma}_{b_{n}}$ are their associated covariance matrices.
The summation assumes that measurement errors are independent across sample indices and uncorrelated between frames.
For notational brevity, we omit the explicit dependence of the residuals and the cost function on the optimization variables throughout this section.

Importantly, the residuals appearing in \eqref{eqn:tls_cost} are not independent: they are coupled through the underlying system model, which imposes constraints between pairs of residuals at each sample index.
These constraints are enforced in the TLS optimization by a set of \emph{condition functions}, one for each observation index $n$.
Starting from the perturbed motion model in \Cref{eq:model_perturbed}, we collect all terms onto one side to obtain the condition function
\begin{equation}
\label{eq:tls_model_perturbed}
\Observation{\Matrix{F}}_{gb_{n}}^{-1}
\Observation{\Matrix{F}}_{ga_{n}}\!
\Matexp{-\Residual{a_{n}}^{\wedge}}
\Prior{\Matrix{F}}_{\nhs\SmallDelta\tau}^{n}
\Matexp{n\,\delta\Vector{\xi}_{\tau}^{\wedge}}
\Prior{\Matrix{F}}_{a_{0}b_{0}}\!
\Matexp{\delta\Vector{\xi}_{ab}^{\wedge}}
\Prior{\Matrix{F}}_{\nhs\SmallDelta\tau}^{-n}
\Matexp{-n\,\delta\Vector{\xi}_{\tau}^{\wedge}}
\Matexp{\Residual{b_{n}}^{\wedge}}
=
\Identity_{5},
\end{equation}
where we have made use of the commutativity of the time-translation subgroup to simplify the expression.
There is one such condition function for each pair of observations at index $n$.

Prior to linearization, we reorder the residual and perturbation terms using the group adjoint in order to express all perturbations in a common tangent space.
Although algebraically tedious due to the multiplicative structure of the group, these steps are shown explicitly to make the subsequent linearization step clear.
Applying a standard adjoint identity yields
\begin{equation}
\label{eq:tls_reordering_1}
\begin{aligned}
\Identity_{5}
& =
\Observation{\Matrix{F}}_{gb_{n}}^{-1}
\Observation{\Matrix{F}}_{ga_{n}}\!
\Matexp{-\Residual{a_{n}}^{\wedge}}
\Prior{\Matrix{F}}_{\nhs\SmallDelta\tau}^{n}
\Matexp{n\,\delta\Vector{\xi}_{\tau}^{\wedge}}
\Prior{\Matrix{F}}_{a_{0}b_{0}}
\Prior{\Matrix{F}}_{\nhs\SmallDelta\tau}^{-n} \\
& \hspace*{1em}
\cdot
\Matexp{
\left(
\LieGroupAdjoint{\Prior{\Matrix{F}}_{\nhs\SmallDelta\tau}^{n}}
\delta\Vector{\xi}_{ab}
\right)^{\wedge}}
\Matexp{-n\,\delta\Vector{\xi}_{\tau}^{\wedge}}
\Matexp{\Residual{b_{n}}^{\wedge}}.
\end{aligned}
\end{equation}
Applying the adjoint again to the time-translation terms gives
\begin{equation}
\label{eq:tls_reordering_2}
\begin{aligned}
\Identity_{5}
& =
\Observation{\Matrix{F}}_{gb_{n}}^{-1}
\Observation{\Matrix{F}}_{ga_{n}}\!
\Matexp{-\Residual{a_{n}}^{\wedge}}
\Prior{\Matrix{F}}_{\nhs\SmallDelta\tau}^{n}
\Prior{\Matrix{F}}_{a_{0}b_{0}}
\Prior{\Matrix{F}}_{\nhs\SmallDelta\tau}^{-n} \\
& \hspace*{1em}
\cdot
\Matexp{
\left(
\LieGroupAdjoint{
\Prior{\Matrix{F}}_{\nhs\SmallDelta\tau}^{n}
\Prior{\Matrix{F}}_{a_{0}b_{0}}^{-1}}
n\,\delta\Vector{\xi}_{\tau}
\right)^{\wedge}}
\Matexp{
\left(
\LieGroupAdjoint{\Prior{\Matrix{F}}_{\nhs\SmallDelta\tau}^{n}}
\delta\Vector{\xi}_{ab}
\right)^{\wedge}}
\Matexp{-n\,\delta\Vector{\xi}_{\tau}^{\wedge}}
\Matexp{\Residual{b_{n}}^{\wedge}}.
\end{aligned}
\end{equation}
Finally, moving the residual $\Residual{a_{n}}\!$ to the right by applying the adjoint once more yields
\begin{equation}
\label{eqn:tls_reordering_3}
\begin{aligned}
\Identity_{5}
& =
\Observation{\Matrix{F}}_{gb_{n}}^{-1}
\Observation{\Matrix{F}}_{ga_{n}}\!
\Prior{\Matrix{F}}_{\nhs\SmallDelta\tau}^{n}
\Prior{\Matrix{F}}_{a_{0}b_{0}}
\Prior{\Matrix{F}}_{\nhs\SmallDelta\tau}^{-n} \\
& \hspace*{1em}
\cdot
\Matexp{
\left(
-\LieGroupAdjoint{
\Prior{\Matrix{F}}_{\nhs\SmallDelta\tau}^{n}
\Prior{\Matrix{F}}_{a_{0}b_{0}}^{-1}
\Prior{\Matrix{F}}_{\nhs\SmallDelta\tau}^{-n}}
\Residual{a_{n}}
\right)^{\wedge}} \\
& \hspace*{1em}
\cdot
\Matexp{
\left(
\LieGroupAdjoint{
\Prior{\Matrix{F}}_{\nhs\SmallDelta\tau}^{n}
\Prior{\Matrix{F}}_{a_{0}b_{0}}^{-1}}
n\,\delta\Vector{\xi}_{\tau}
\right)^{\wedge}}
\Matexp{
\left(
\LieGroupAdjoint{
\Prior{\Matrix{F}}_{\nhs\SmallDelta\tau}^{n}}
\delta\Vector{\xi}_{ab}
\right)^{\wedge}}
\Matexp{-n\,\delta\Vector{\xi}_{\tau}^{\wedge}}
\Matexp{\Residual{b_{n}}^{\wedge}}.
\end{aligned}
\end{equation}

To linearize the condition functions, we take the logarithm of \eqref{eqn:tls_reordering_3} and apply the BCH formula repeatedly, retaining only first-order terms in the perturbations and discarding higher-order commutator terms.
Under this approximation, products of exponentials reduce to additive expressions in the Lie algebra.
For compactness, we first define the predicted relative transformation at index $n$ as
\begin{equation}
\Prior{\Matrix{F}}_{a_{n}b_{n}}
\Defined
\Prior{\Matrix{F}}_{\nhs\SmallDelta\tau}^{n}
\Prior{\Matrix{F}}_{a_{0}b_{0}}
\Prior{\Matrix{F}}_{\nhs\SmallDelta\tau}^{-n}.
\end{equation}
The error at the current linearization point is
\begin{equation}
\Vector{\varepsilon}_{n}
=
\Matlog{
\Observation{\Matrix{F}}_{gb_{n}}^{-1}
\Observation{\Matrix{F}}_{ga_{n}}\!
\Prior{\Matrix{F}}_{a_{n}b_{n}}
}^{\vee},
\end{equation}
and each linearized condition function takes the form
\begin{equation}
\Vector{\varepsilon}_{n}
+
\Matrix{\mathcal{J}}_{a_{n}} \Residual{a_{n}}
+
\Matrix{\mathcal{J}}_{b_{n}} \Residual{b_{n}}
+
\Matrix{\mathcal{J}}_{ab,n}\,\delta\Vector{\xi}_{ab}
+
\Matrix{\mathcal{J}}_{\tau,n}\,\delta\tau
\approx
\Vector{0}.
\end{equation}
Here, the Jacobian matrices $\Matrix{\mathcal{J}}_{(\cdot)}$ and $\Matrix{\mathcal{J}}_{(\cdot),n}$ are obtained by linearizing the logarithm of the condition function with respect to the perturbation variables.
For compactness, we use subscripted identifiers to indicate the variable with respect to which each Jacobian is taken, rather than introducing longer derivative notation.
The explicit dependence on the sample index $n$ reflects the fact that the predicted transformation $\Prior{\Matrix{F}}_{a_{n}b_{n}}$ varies with $n$, while the perturbation vector $\delta\Vector{\mathcal{X}}$ is shared across all samples.

The Jacobians with respect to the observation residuals are
\begin{align}
\Matrix{\mathcal{J}}_{a_{n}}
& =
-
\RightJacobianSGal(\Vector{\varepsilon}_{n})^{-1}
\LieGroupAdjoint{\Prior{\Matrix{F}}_{a_{n}b_{n}}^{-1}}, \\
\Matrix{\mathcal{J}}_{b_{n}}
& =
\RightJacobianSGal(\Vector{\varepsilon}_{n})^{-1}.
\end{align}
The Jacobians with respect to the model parameters are
\begin{align}
\Matrix{\mathcal{J}}_{\tau,n}
& =
\RightJacobianSGal(\Vector{\varepsilon}_{n})^{-1}
\left(
n\,\LieGroupAdjoint{
\Prior{\Matrix{F}}_{\nhs\SmallDelta\tau}^{n}
\Prior{\Matrix{F}}_{a_{0}b_{0}}^{-1}}
-
n\,\Identity_{10}
\right)
\Vector{e}_{10}, \\
\Matrix{\mathcal{J}}_{ab,n}
& =
\RightJacobianSGal(\Vector{\varepsilon}_{n})^{-1}
\LieGroupAdjoint{\Prior{\Matrix{F}}_{\nhs\SmallDelta\tau}^{n}},
\end{align}
Here, $\RightJacobianSGal(\cdot)$ denotes the right Jacobian of the $\LieGroupSGal{3}$ exponential map, and its inverse appears as a result of linearizing the logarithm in \Cref{eqn:tls_cost} according to \Cref{eqn:bch_right_jacobian}.
The vector $\Vector{e}_{10}$ is the 10th canonical basis vector, selecting the column of the Jacobian corresponding to the time perturbation $\delta\tau$.

To solve the TLS problem, we eliminate the residual variables using Lagrange multipliers, yielding an equivalent reduced system in the parameter perturbations.
This procedure results in an effective covariance that captures how observation uncertainty propagates through the linearized condition functions.
Defining the effective covariance of the linearized condition function as
\begin{equation}
\Matrix{\Sigma}_{n}
=
\Matrix{\mathcal{J}}_{a_{n}}
\Matrix{\Sigma}_{a_{n}}
\Matrix{\mathcal{J}}_{a_{n}}^{\T}
+
\Matrix{\mathcal{J}}_{b_{n}}
\Matrix{\Sigma}_{b_{n}}
\Matrix{\mathcal{J}}_{b_{n}}^{\T},
\end{equation}
the TLS normal equations take the form
\begin{equation}
\left(
\sum_{n = 0}^{N}
\Matrix{\mathcal{J}}_{\SmallStateVector,n}^{\T}
\Matrix{\Sigma}_{n}^{-1}
\Matrix{\mathcal{J}}_{\SmallStateVector,n}
\right)
\delta\Vector{\mathcal{X}}
=
-
\sum_{n = 0}^{N}
\Matrix{\mathcal{J}}_{\SmallStateVector,n}^{\T}
\Matrix{\Sigma}_{n}^{-1}
\Vector{\varepsilon}_{n},
\end{equation}
where
$\Matrix{\mathcal{J}}_{\SmallStateVector,n}
=
\bbm
\Matrix{\mathcal{J}}_{ab,n}\nds & \Matrix{\mathcal{J}}_{\tau,n}
\ebm
\in \Real^{10 \times 11}$.

Solving this system yields the optimal perturbation $\delta\Vector{\mathcal{X}}$, which is applied on the group via
\begin{align}
\Estimate{\Matrix{F}}_{a_{0}b_{0}}
&\leftarrow
\Prior{\Matrix{F}}_{a_{0}b_{0}}
\Matexp{\delta\Vector{\xi}_{ab}^{\wedge}}, \\
\SmallDelta\Estimate{\tau}
&\leftarrow
\SmallDelta\Prior{\tau} + \delta\tau.
\end{align}
Here, the $(\Estimate{\cdot})$ denotes an updated estimate obtained after applying the computed perturbation.
This updated estimate then serves as the linearization point for the subsequent iteration.
The procedure is iterated until convergence.

\subsection{Simulation studies}
\label{subsec:simulations}

We evaluate the proposed estimation framework through Monte Carlo simulation studies based on the problem described in the preceding sections.
In each trial, two inertial reference frames move apart along distinct worldlines, and noisy observations of position, orientation, velocity, and local time are generated along each trajectory. 
See \Cref{fig:evolving_frames} for a visual representation of the problem.

We compare the performance of two estimator formulations that differ in how they model time and its interaction with spatial quantities.
The first formulation operates directly on the special Galilean group $\LieGroupSGal{3}$, treating both the initial relative transformation between frames and the inter-measurement time increment as group elements.
In this case, all operations, perturbations, and uncertainties are represented on the group and propagated in its tangent space, fully capturing the intrinsic coupling between position, orientation, velocity, and time.

The second formulation is based on the group of extended poses $\LieGroupSETwo{3}$ \cite{2014_Barrau_Invariant}, which augments rigid-body transformations with velocity but does not include time as a group component.
Here, temporal quantities---including the initial time offset between frames and the inter-measurement interval---are treated as independent Euclidean variables, resulting in a direct product structure between $\LieGroupSETwo{3}$ and $\Real$.
Because time is not embedded in the group action, temporal uncertainty does not propagate into spatial uncertainty through the geometry of the group.

In the simulations, each trial spans a total duration of 100~s, with measurements collected at 1~s intervals, resulting in 101 measurements along each worldline for frames $\CoordinateFrame{a}$ and $\CoordinateFrame{b}$.
Throughout all simulations, measurement noise is assumed to be zero-mean and Gaussian with diagonal covariance.
The observation noise covariance for each frame, ordered by position, velocity, orientation, and time components, is given by
\begin{equation}
\Matrix{\Sigma}_{a_{n}} 
=
\Matrix{\Sigma}_{b_{n}}
=
\mathrm{diag}
\left(
\bbm
0.2^2 &\nds\nhs
0.2^2 &\nds\nhs
0.2^2 &\nds\nhs
0.2^2 &\nds\nhs
0.2^2 &\nds\nhs
0.2^2 &\nds\nhs
(\pi/60)^2 &\nds\nhs
(\pi/60)^2 &\nds\nhs
(\pi/60)^2 &\nds\nhs
0.2^2
\ebm
\right).
\end{equation}
Additional zero-mean Gaussian noise is applied to the inter-measurement time increment $\SmallDelta\tau$, with variance $\sigma_{\SmallDelta\tau}^{2} = 0.2^2$.

In the following, we describe the metrics used to evaluate estimator accuracy and consistency, present the results of the Monte Carlo simulations, and discuss the implications of the observed differences between these formulations.

\begin{table}
\renewcommand{\arraystretch}{1.1}
\centering
\label{tab:simulation}
\begin{tabular}{rrcc}
\toprule
\multicolumn{2}{c}{Evaluation Metric} &
$\LieGroupSGal{3}$ &
$\LieGroupSETwo{3} \times \Real$ \\
\midrule
\multirow{5}{3.4em}{\centering RMSE}
& Position ($\Vector{r}_{a_{0}b_{0}}$)       & 0.441 & 0.575 \\
& Velocity ($\Vector{v}_{a_{0}b_{0}}$)       & 0.069 & 0.071 \\
& Orientation ($\Vector{\phi}_{a_{0}b_{0}}$) & 0.013 & 0.013 \\
& Time Offset ($\tau_{a_{0}b_{0}}$)          & 0.027 & 0.028 \\
& Interval ($\Delta\tau$)                    & 0.004 & 0.004 \\
\midrule
\multicolumn{2}{r}{ANEES} & 12.49 & 97.08 \\
\multicolumn{2}{r}{MMD}   & 3.45  & 8.99  \\
\bottomrule
\end{tabular}
\caption{Performance comparison of three estimator formulations over 1,200 Monte Carlo trials.
Results are shown for estimation on $\LieGroupSGal{3}$ and on $\LieGroupSETwo{3}$ with time treated independently, that is, $\LieGroupSETwo{3} \times \Real$.
RMSE is reported separately for each estimated component of the initial relative transformation and timing parameters.}
\end{table}

\subsubsection{Metrics}

Estimator performance is evaluated using a combination of accuracy and consistency metrics computed over $N$ Monte Carlo trials.
For each trial $k$, let $\hat{\mathbf{X}}_k$ denote the estimated group element, $\mathbf{X}_k$ the ground truth, and
\begin{equation}
\Vector{\varepsilon}_{k} 
=
\Matlog{\hat{\mathbf{X}}_{k}^{-1}\mathbf{X}_{k}}
\in \mathbb{R}^d
\end{equation}
the corresponding estimation error expressed in the Lie algebra, where $d$ denotes the dimension of the appropriate parameter vector (and with a slight abuse of our notation from the previous section).
Let $\Matrix{P}_{k} \in \Real^{d \times d}$ denote the estimated covariance associated with $\hat{\Matrix{X}}_{k}$.
The covariance $\Matrix{P}_k$ is approximated using a Laplace approximation about the converged solution, obtained as the inverse of the Gauss-Newton Hessian.

We first quantify estimation accuracy using the root-mean-square error (RMSE),
\begin{equation}
\mathrm{RMSE}
=
\sqrt{\frac{1}{N}\sum_{k=1}^N \lVert \Vector{\varepsilon}_k \rVert^2},
\end{equation}
which measures the average magnitude of the estimation error over all trials.
To avoid combining quantities with incompatible physical units, RMSE is reported on a per-component basis only (i.e., position, velocity, orientation, and time).

To account for estimator uncertainty, we also report the mean Mahalanobis distance (MMD),
\begin{equation}
\mathrm{MMD}
=
\frac{1}{N}\sum_{k=1}^N
\sqrt{\Vector{\varepsilon}_k^{\T} \Matrix{P}_k^{-1}\Vector{\varepsilon}_k}.
\end{equation}
This dimensionless metric measures the average estimation error normalized by the reported covariance and can be interpreted as the expected number of standard deviations separating the estimate from the true state.
While MMD provides a compact scalar summary of normalized accuracy, it does not by itself assess statistical consistency.

Estimator consistency is evaluated using the normalized estimation error squared (NEES),
\begin{equation}
\epsilon_k
=
\Vector{\varepsilon}_k^\top \Matrix{P}_k^{-1}\Vector{\varepsilon}_k,
\end{equation}
and its average over the Monte Carlo trials,
\begin{equation}
\mathrm{ANEES}
 =
\frac{1}{N}\sum_{k=1}^N \epsilon_k.
\end{equation}
For a statistically consistent estimator, $\epsilon_k$ follows a $\chi^2_d$ distribution, and the sum $\sum_{k=1}^N \epsilon_k$ follows a $\chi^2_{Nd}$ distribution.
Consistency is therefore assessed by comparing the empirical ANEES value against the corresponding confidence interval derived from the $\chi^2$ distribution.

Together, RMSE characterizes absolute estimation accuracy, MMD summarizes accuracy relative to the reported uncertainty, and ANEES evaluates the statistical consistency of the estimator covariance.

\subsubsection{Results}

Across the Monte Carlo trials, both estimator formulations achieve broadly comparable accuracy across all estimated components, with the Galilean formulation consistently matching or slightly outperforming the decoupled model.
In particular, estimation on $\LieGroupSGal{3}$ yields lower RMSE for position (0.441 vs.\ 0.575) and velocity (0.069 vs.\ 0.071), while orientation, time offset, and inter-measurement estimation errors are essentially the same across formulations.
These results indicate that, in terms of point accuracy alone, the two approaches perform similarly, with only modest gains attributable to the Galilean formulation.

A substantially different picture emerges, however, when the statistical consistency of the estimators is examined.
The Galilean estimator exhibits an ANEES of 12.49, close to the expected value of 11 for an 11-parameter estimate, but just slightly above the corresponding 95\% $\chi^2$ confidence interval computed over 1,200 Monte Carlo trials ($10.74 \leq \mathrm{ANEES} \leq 11.27$).
In contrast, the decoupled $\LieGroupSETwo{3}\times\Real$ formulation produces an ANEES of 97.08, demonstrating very significant overconfidence despite the comparable RMSE.
This discrepancy is further reflected in the mean Mahalanobis distance (MMD), which is 3.45 for $\LieGroupSGal{3}$, close to the theoretical expectation of approximately 3.24, compared to 8.99 for the decoupled formulation.
Together, these findings show that while both methods achieve similar accuracy, only the Galilean formulation yields statistically consistent estimates by accounting for the coupled interactions among position, orientation, velocity, and time.
This suggests an advantage to performing estimation directly on $\LieGroupSGal{3}$ in applications where reliable uncertainty quantification is critical.

\section{Conclusion}
\label{sec:conclusion}

In this paper, we have developed a geometric formulation of uncertainty on the special Galilean group, motivated by problems in which time, velocity, and spatial configuration are fundamentally coupled.
We provided a comprehensive overview of the group structure and its associated Lie-theoretic machinery.
By embedding time directly within the group structure, $\LieGroupSGal{3}$ captures the intrinsic coupling between temporal uncertainty, relative velocity, and spatial displacement that arises in physically consistent inertial motion and the symmetries of Galilean spacetime.
The Lie group $\LieGroupSGal{3}$ provides a natural setting for these problems and for treating the associated uncertainty; this short report provides some of the necessary mathematical machinery.

To illustrate the utility of the group operations under uncertainty we introduced an estimation problem in which inertial frames evolve relative to one another over time.
We compared batch estimators based on the full $\LieGroupSGal{3}$ model against formulations that either decouple time as an independent scalar or neglect time uncertainty altogether.
Through Monte Carlo simulation studies, we compared batch estimators operating on $\LieGroupSGal{3}$ with an estimator based on a decoupled formulation using $\LieGroupSETwo{3}\times\mathbb{R}$.
While both formulations produced comparable point estimates in terms of root-mean-square error, the Galilean formulation exhibited markedly improved statistical consistency.
In particular, the average normalized estimation error squared (ANEES) for $\LieGroupSGal{3}$ closely matched the theoretical $\chi^2$ bounds, whereas the decoupled formulation was systematically overconfident.
This behavior arises directly from the group structure: in $\LieGroupSGal{3}$, uncertainty in time propagates naturally into spatial uncertainty through velocity, whereas this coupling is absent in direct-product formulations.

Our results show that the benefits of the Galilean group are not merely representational, but statistical: by respecting the semidirect coupling between time, velocity, and position, $\LieGroupSGal{3}$ induces uncertainty geometries that align with the underlying physics of inertial motion. 
This leads to estimators that are not only accurate, but also properly calibrated.
More broadly, our work highlights the importance of selecting state spaces whose group structure faithfully reflects the kinematic and temporal symmetries of the problem at hand, particularly in estimation settings where consistency is as critical as accuracy.

We expect these ideas to be relevant across a range of applications, including multi-sensor fusion, inertial navigation, and collaborative or distributed autonomous systems operating over extended spatial and temporal scales.

\setcounter{section}{0}
\renewcommand{\thesection}{A}
\section*{Appendix}
\label{sec:appendix}

For completeness, this appendix collects several identities useful for manipulating expressions involving products of skew-symmetric matrices.
The first identity is known and is included here for reference, while the remaining two identities appear to be new, to the best of the authors' knowledge.
These identities are used to derive the compact form of the Jacobian in~\Cref{eqn:SGal3_left_jacobian_definition} of \Cref{sec:SGal_group}.

\begin{identity}
\label{ident:skew_negative}
Let $\Vector{u} \in \Real^{3}$ and let $\Norm{\Vector{u}} = 1$. Then the following identity holds:
\begin{equation}
\label{eqn:identity_1}
\uw\uw\uw = -\uw.
\end{equation}
\end{identity}

\begin{proof}
The result follows by direct expansion and manipulation; full details are given in~\cite{2023_Kelly_Galilean}.
\end{proof}

\begin{identity}
\label{ident:skew_products_triple}
Let $\Vector{u}, \Vector{v} \in \Real^{3}$ and let $\Norm{\Vector{u}} = 1$. Then following identity holds:
\begin{equation}
\label{eqn:identity_2}
\uw\Vector{v}^{\wedge}\uw - 
\uw\uw\Vector{v}^{\wedge} -
\Vector{v}^{\wedge}\uw\uw 
= 
\Vector{v}^{\wedge}.
\end{equation}
\end{identity}

\begin{proof}
The result follows by direct expansion and manipulation; full details are given in~\cite{2023_Kelly_Galilean}.
\end{proof}

\begin{identity}
\label{ident:skew_products_quad}
Let $\Vector{u}, \Vector{v} \in \Real^{3}$ and let $\Norm{\Vector{u}} = 1$. Then the following identity holds:
\begin{equation}
\label{eqn:identity_3}
\uw\uw\Vector{v}^{\wedge}\uw
=
\uw\Vector{v}^{\wedge}\uw\uw.
\end{equation}
\end{identity}

\begin{proof}
Rearranging \Cref{ident:skew_products_triple}, right-multiplying by $\uw$, and making use of \Cref{ident:skew_negative} yields
\begin{align*}
\uw\uw\Vector{v}^{\wedge}\uw 
& =
\left(
\uw\Vector{v}^{\wedge}\uw - 
\Vector{v}^{\wedge}\uw\uw -
\Vector{v}^{\wedge}
\right)
\uw \\[1mm]
& =
\uw\Vector{v}^{\wedge}\uw\uw - 
\Vector{v}^{\wedge}\uw\uw\uw -
\Vector{v}^{\wedge}\uw \\
& =
\uw\Vector{v}^{\wedge}\uw\uw + 
\Vector{v}^{\wedge}\uw -
\Vector{v}^{\wedge}\uw \\
& =
\uw\Vector{v}^{\wedge}\uw\uw.
\qedhere
\end{align*}
\end{proof}

\bibliography{2025-kelly-uncertainty-rsta.bib}

\begin{thebibliography}{25}
\providecommand{\natexlab}[1]{#1}
\providecommand{\url}[1]{\texttt{#1}}
\expandafter\ifx\csname urlstyle\endcsname\relax
  \providecommand{\doi}[1]{doi: #1}\else
  \providecommand{\doi}{doi: \begingroup \urlstyle{rm}\Url}\fi

\bibitem[Arnold(1989)]{1989_Arnold_Mathematical}
Vladimir~Igorevich Arnold.
\newblock \emph{Mathematical Methods of Classical Mechanics}, volume~60 of \emph{Graduate Texts in Mathematics}.
\newblock Springer-Verlag, New York, New York, USA, 2nd edition, 1989.
\newblock \doi{10.1007/978-1-4757-1693-1}.

\bibitem[Artz(1981)]{1981_Artz_Classical}
Ray~E. Artz.
\newblock Classical mechanics in {Galilean} space-time.
\newblock \emph{Foundations of Physics}, 11\penalty0 (9):\penalty0 679--697, October 1981.
\newblock \doi{10.1007/BF00726944}.

\bibitem[Barfoot(2024)]{2024_Barfoot_State}
Timothy~D. Barfoot.
\newblock \emph{State Estimation for Robotics}.
\newblock Cambridge University Press, New York, New York, USA, 2nd edition, 2024.

\bibitem[Barfoot(2025)]{2025_Barfoot_Integral}
Timothy~D. Barfoot.
\newblock Integral forms in matrix {Lie} groups.
\newblock arXiv preprint arXiv:2503.02820, 2025.

\bibitem[Barfoot and Furgale(2014)]{2014_Barfoot_Associating}
Timothy~D. Barfoot and Paul~T. Furgale.
\newblock Associating uncertainty with three-dimensional poses for use in estimation problems.
\newblock \emph{{IEEE} Transactions on Robotics}, 30\penalty0 (3):\penalty0 679--693, June 2014.
\newblock \doi{10.1109/TRO.2014.2298059}.

\bibitem[Barrau and Bonnabel(2014)]{2014_Barrau_Invariant}
Axel Barrau and Silv\`{e}re Bonnabel.
\newblock Invariant particle filtering with application to localization.
\newblock In \emph{Proceedings of the {IEEE} Conference on Decision and Control {(CDC)}}, pages 5599--5605, December 2014.
\newblock \doi{10.1109/CDC.2014.7040265}.

\bibitem[Barrau and Bonnabel(2023)]{2023_Barrau_Geometry}
Axel Barrau and Silv\`{e}re Bonnabel.
\newblock The geometry of navigation problems.
\newblock \emph{{IEEE} Transactions on Automatic Control}, 68\penalty0 (2):\penalty0 689--704, February 2023.
\newblock \doi{10.1109/TAC.2022.3144328}.

\bibitem[Bhand and Lewis(2005)]{2005_Bhand_Rigid}
Ajit Bhand and Andrew~D. Lewis.
\newblock Rigid body mechanics in {Galilean} spacetimes.
\newblock \emph{Journal of Mathematical Physics}, 46\penalty0 (10):\penalty0 102902, October 2005.
\newblock \doi{10.1063/1.2060547}.

\bibitem[Brossard et~al.(2022)Brossard, Barrau, Chauchat, and Bonnabel]{2022_Brossard_Associating}
Martin Brossard, Axel Barrau, Paul Chauchat, and Silv\`{e}re Bonnabel.
\newblock Associating uncertainty to extended poses for on {Lie} group {IMU} preintegration with rotating earth.
\newblock \emph{{IEEE} Transactions on Robotics}, 38\penalty0 (2):\penalty0 998--1015, April 2022.
\newblock \doi{10.1109/TRO.2021.3100156}.

\bibitem[Chirikjian(2011)]{2011_Chirikjian_Stochastic}
Gregory~S. Chirikjian.
\newblock \emph{Stochastic Models, Information Theory, and Lie Groups, Volume 2: Analytic Methods and Modern Applications}.
\newblock Applied and Numerical Harmonic Analysis. Birkh\"{a}user Basel, 2011.
\newblock \doi{10.1007/978-0-8176-4944-9}.

\bibitem[Delama et~al.(2025)Delama, Fornasier, Mahony, and Weiss]{2025_Delama_Equivariant}
Giulio Delama, Alessandro Fornasier, Robert Mahony, and Stephan Weiss.
\newblock Equivariant {IMU} preintegration with biases: A {Galilean} group approach.
\newblock \emph{{IEEE} Robotics and Automation Letters}, 10\penalty0 (1):\penalty0 724--731, January 2025.
\newblock \doi{10.1109/LRA.2024.3511424}.

\bibitem[Giefer(2021)]{2021_Giefer_Uncertainties}
Lino~Antoni Giefer.
\newblock Uncertainties in {Galilean} spacetime.
\newblock In \emph{Proceedings of the {IEEE} International Conference on Information Fusion {(FUSION)}}, Sun City, South Africa, November 2021.
\newblock \doi{10.23919/FUSION49465.2021.9627044}.

\bibitem[Holm(2011)]{2011_Holm_Geometric_Part_II}
Daryl~D. Holm.
\newblock \emph{Geometric Mechanics - Part II: Rotating, Translating and Rolling}.
\newblock Imperial College Press, London, United Kingdom, 2 edition, November 2011.
\newblock \doi{10.1142/p802}.

\bibitem[Kelly(2023)]{2023_Kelly_Galilean}
Jonathan Kelly.
\newblock All about the {Galilean} group $\mathrm{SGal}(3)$.
\newblock Technical Report STARS-2023-001, University of Toronto, Toronto, Ontario, Canada, November 2023.
\newblock doi: 10.48550/arXiv.2312.07555.

\bibitem[Kelly and Sukhatme(2011)]{2011_Kelly_Visual}
Jonathan Kelly and Gaurav~S. Sukhatme.
\newblock Visual-inertial sensor fusion: Localization, mapping and sensor-to-sensor self-calibration.
\newblock \emph{The International Journal of Robotics Research}, 30\penalty0 (1):\penalty0 56--79, January 2011.
\newblock \doi{10.1177/0278364910382802}.

\bibitem[L\'{e}vy-Leblond(1971)]{1971_Levy-Leblond_Galilei}
Jean-Marc L\'{e}vy-Leblond.
\newblock {Galilei} group and {Galilean} invariance.
\newblock In Ernest~M. Loebl, editor, \emph{Group Theory and its Applications}, volume~II, pages 221--299. Academic Press, 1971.
\newblock \doi{10.1016/B978-0-12-455152-7.50011-2}.

\bibitem[Long et~al.(2012)Long, Wolfe, Mashner, and Chirikjian]{2012_Long_Banana}
Andrew Long, Kevin Wolfe, Michael Mashner, and Gregory Chirikjian.
\newblock The banana distribution is {Gaussian}: A localization study with exponential coordinates.
\newblock In \emph{Proceedings of Robotics: Science and Systems {(RSS)}}, Sydney, Australia, July 2012.
\newblock \doi{10.15607/RSS.2012.VIII.034}.

\bibitem[Mahony et~al.(2025)Mahony, Kelly, and Weiss]{2025_Mahony_Galilean}
Robert Mahony, Jonathan Kelly, and Stephan Weiss.
\newblock {Galilean} symmetry in robotics.
\newblock arXiv preprint arXiv:2510.10468, 2025.

\bibitem[Marsden and Ratiu(1999)]{1999_Marsden_Introduction}
Jerrold~E. Marsden and Tudor~S. Ratiu.
\newblock \emph{Introduction to Mechanics and Symmetry: A Basic Exposition of Classical Mechanical Systems}, volume~17 of \emph{Texts in Applied Mathematics}.
\newblock Springer, New York, New York, USA, 2nd edition, April 1999.
\newblock \doi{10.1007/978-0-387-21792-5}.

\bibitem[Maudlin(2012)]{2012_Maudlin_Philosophy}
Tim Maudlin.
\newblock \emph{Philosophy of Physics: Space and Time}, volume~5 of \emph{Princeton Foundations of Contemporary Philosophy}.
\newblock Princeton University Press, Princeton, New Jersey, USA, 2012.

\bibitem[Murray et~al.(1994)Murray, Li, and Sastry]{1994_Murray_Mathematical}
Richard~M. Murray, Zexiang Li, and S.~Shankar Sastry.
\newblock \emph{A Mathematical Introduction to Robotic Manipulation}.
\newblock CRC Press, Inc., Boca Raton, Florida, USA, 1st edition, 1994.

\bibitem[Penrose(2005)]{2005_Penrose_Road}
Roger Penrose.
\newblock \emph{The Road to Reality: A Complete Guide to the Laws of the Universe}.
\newblock Alfred A. Knopf, London, United Kingdom, 2005.

\bibitem[Selig(2005)]{2005_Selig_Geometric}
Jon~M. Selig.
\newblock \emph{Geometric Fundamentals of Robotics}.
\newblock Monographs in Computer Science. Springer-Verlag, New York, New York, USA, 2nd edition, 2005.
\newblock \doi{10.1007/b138859}.

\bibitem[Sol\`{a} et~al.(2021)Sol\`{a}, Deray, and Atchuthan]{2021_Sola_Micro}
Joan Sol\`{a}, Jeremie Deray, and Dinesh Atchuthan.
\newblock A micro {Lie} theory for state estimation in robotics.
\newblock arXiv preprint arXiv:1812.01537, 2021.

\bibitem[Wolfe et~al.(2011)Wolfe, Mashner, and Chirikjian]{2011_Wolfe_Bayesian}
Kevin Wolfe, Michael Mashner, and Gregory Chirikjian.
\newblock {Bayesian} fusion on {Lie} groups.
\newblock \emph{Journal of Algebraic Statistics}, 2\penalty0 (1):\penalty0 75--97, 2011.
\newblock \doi{10.18409/jas.v2i1.11}.

\end{thebibliography}

\end{document}